\documentclass[apj]{emulateapj}

\usepackage{times}
\usepackage{amssymb}
\usepackage{amsmath}
\usepackage{pifont}
\usepackage{graphics}
\usepackage[svgnames]{xcolor}
\usepackage{epsfig}
\usepackage{threeparttable}

\usepackage[normalem]{ulem}
\usepackage{color}

\newcommand{\MS}{M\textsubscript{$\odot$}}    
\newcommand{\RS}{R\textsubscript{$\odot$}}    
\newcommand{\RP}{R\textsubscript{P}}          
\newcommand{\ME}{M\textsubscript{$\oplus$}}   
\newcommand{\RE}{R\textsubscript{$\oplus$}}   
\newcommand{\Mstar}{M\textsubscript{$\star$}} 
\newcommand{\Rstar}{R\textsubscript{$\star$}} 

\newcommand{\bjdtdb}{\ensuremath{\rm {BJD_{TDB}}}}

\begin{document}


\title{A Multi-Year Search For Transits Of Proxima Centauri.\\I: Light Curves Corresponding To Published Ephemerides.}

\author{David L. Blank\altaffilmark{1}, Dax Feliz\altaffilmark{2,3,4}, Karen A. Collins\altaffilmark{5}, Graeme L. White\altaffilmark{1}, Keivan G.\ Stassun\altaffilmark{2,3}, Ivan A. Curtis\altaffilmark{6}, Rhodes Hart\altaffilmark{1}, John F. Kielkopf\altaffilmark{7}, Peter Nelson\altaffilmark{8}, Howard Relles\altaffilmark{5},
Christopher Stockdale\altaffilmark{9}, Bandupriya Jayawardene\altaffilmark{10}, Carlton R. Pennypacker\altaffilmark{11}, Paul Shankland\altaffilmark{12}, Daniel E. Reichart\altaffilmark{13}, Joshua B. Haislip\altaffilmark{13}, and Vladimir V. Kouprianov\altaffilmark{13}}

\altaffiltext{1}{University of Southern Queensland, Computational Engineering and Science Research Centre, Toowoomba, Queensland 4350, Australia}
\altaffiltext{2}{Department of Physics, Fisk University, 1000 17th Avenue North, Nashville, TN 37208, USA}
\altaffiltext{3}{Department of Physics and Astronomy, Vanderbilt University, Nashville, TN 37235, USA}
\altaffiltext{4}{Corresponding Author, dax.feliz@vanderbilt.edu}
\altaffiltext{5}{Harvard-Smithsonian Center for Astrophysics, Cambridge, MA 02138, USA}
\altaffiltext{6}{ICO, Adelaide, South Australia}
\altaffiltext{7}{Department of Physics and Astronomy, University of Louisville, Louisville, KY 40292, USA}
\altaffiltext{8}{Ellinbank Observatory, Victoria, Australia}
\altaffiltext{9}{Hazelwood Observatory, Churchill, Victoria, Australia}
\altaffiltext{10}{Physical Sciences Group, College of Science and Engineering, James Cook University, Townsville, QLD 4811, Australia}
\altaffiltext{11}{Lawrence Berkeley National Laboratory and Dept. of Physics, University of California Berkeley, Berkeley CA 94720-7300, USA}
\altaffiltext{12}{U.S. Naval Observatory, Flagstaff Station, 10391 W Naval Observatory Rd, Flagstaff, AZ 86001, USA}
\altaffiltext{13}{Department of Physics and Astronomy, University of North Carolina at Chapel Hill, Campus Box 3255, Chapel Hill, NC 27599, USA}



\begin{abstract}

Proxima Centauri has become the subject of intense study since the radial-velocity discovery by \citet{Anglada:2016} of a planet orbiting this nearby M-dwarf every $\sim11.2$~days. If Proxima Centauri~b transits its host star, independent confirmation of its existence is possible, and its mass and radius can be measured in units of the stellar host mass and radius. To date, there have been three independent claims of possible transit-like event detections in light curve observations obtained by the MOST satellite (in 2014--15), the BSST telescope in Antarctica (in 2016), and the Las Campanas Observatory (in 2016). The claimed possible detections are tentative, due in part to the variability intrinsic to the host star, and in the case of the ground-based observations, also due to the limited duration of the light curve observations. Here, we present preliminary results from an extensive photometric monitoring campaign of Proxima Centauri, using telescopes around the globe and spanning from 2006 to 2017, comprising a total of 329 observations. Considering our data that coincide directly and/or phased with the previously published tentative transit detections, we are unable to independently verify those claims. We do, however, verify the previously reported ubiquitous and complex variability of the host star. We discuss possible interpretations of the data in light of the previous claims, and we discuss future analyses of these data that could more definitively verify or refute the presence of transits associated with the radial-velocity discovered planet.
\end{abstract}

\section{Introduction}\label{sec:intro}

 Neptune-like and lower mass planets are common around M dwarfs; a result predicted in simulations \citep{Laughlin:2004,Ida:2004,Montgomery:2009}, validated within a factor of $\sin i$ in radial-velocity (RV) observations (e.g. \citealt{Bonfils:2005,Bonfils:2013}), and now confirmed by Kepler transit work \citep{Dressing:2013,Dressing:2015, Morton:2014}. This is of great interest since M dwarfs far outnumber other stellar types and terrestrial analogs with orbital periods of only a few days to a few weeks could still have an Earth-like climate, despite being tidally locked \citep{Joshi:1997} or subject to high UV flux \citep{France:2013}, though such factors make it almost certain that the exoplanet history will be very different than that of Earth \citep{Lopez:2012,Ramirez:2014,Luger:2015}. Such planets would be prime candidates in searches for spectroscopic evidence of life by space-born missions (see \citealt{Tarter:2007} and \citealt{Shields:2016} for useful reviews).

The discovery of Proxima Centauri~b (Proxima~b, hereafter) claimed by \citet{Anglada:2016} (A2016, hereafter) in an 11.2~day habitable zone \citep{Kopparapu:2013} orbit of its host star is important because the planet would likely be a rocky \citep{Brugger:2016,Kane:2017, Bixel:2017} and possibly habitable world \citep{Ribas:2016, Barnes:2016,Meadows:2016,Turbet:2016,Boutle:2017} orbiting our nearest-known stellar neighbor. The main source for the uncertainty in the nature and habitability of Proxima~b is that only its lower mass limit of 1.27 M$_{\oplus}$ (A2016) is reported since it was discovered using the RV technique. If Proxima~b transits its host star, independent confirmation of its existence is possible, and its mass and radius can be measured in units of the stellar host mass and radius. Thus, it would be possible to infer bulk composition \citep {Lopez:2014,Weiss:2014, Rogers:2015,Chen:2017}, which in turn impacts the question of habitability. \citet{Damasso:2017} (D2017, hereafter) reanalyzed the RV data and provided a refined ephemeris (see Section \ref{sec:Damasso2017}).

There have been searches for transiting planets orbiting Proxima Centauri (Proxima, hereafter) from both space-based \citep{Kipping:2017} and ground-based \citep{Liu:2017,Li:2017} observatories, but no convincing transit candidates were found. However, \citet{Kipping:2017} (K2017, hereafter) and \citet{Liu:2017} (L2017, hereafter) describe tentative transit-like event detections compatible with the RV orbit, and K2017 and \citet{Li:2017} (Li2017, hereafter) describe potential transit-like detections that are incompatible with the RV orbit.

We have been conducting a search for transiting planets orbiting Proxima since 2006 as part of our Global Earth M-dwarf Search Survey (GEMSS)\footnote{https://gemss.wordpress.com}. As was pointed out in \citet{Shankland:2006} and \citet{Nutzman:2008}, sub-meter diameter telescopes with commercial grade CCD cameras provide sufficient photometric precision to detect transiting terrestrial-type exoplanets around mid- and late-M dwarfs. For example, Proxima~b is predicted to have a radius between 0.9 and 1.4 R$_{\oplus}$ \citep{Brugger:2016}, which would produce a detectable transit with a depth in the range of $0.5-1.3\%$. The observations from the first year of GEMSS operation are described in \citet{Blank:2007}\footnote{Available at https://gemss.files.wordpress.com/2007/04/gemss2.jpg}.

The observations reported in Section \ref{sec:observations} are the basis of two studies. In this work, we examine the question of whether Proxima~b transits, specifically in the context of the three published RV ephemerides (A2016, K2017, D2017) and in the context of tentative transit-like detection claims by K2017, L2017 and Li2017. Section~\ref{sec:previous} summarizes the recent claims in the literature that we specifically examine in this work. Section~\ref{sec:observations} presents our new observations and data reduction methodology. The results, discussion, and summary of conclusions are provided in Sections \ref{sec:results}, \ref{sec:discussion}, and \ref{sec:conclusion}, respectively. 
 
In D.F. et al. 2018, in preparation (Paper~II, hereafter), we will present a more general transit search based on our Proxima data, including a search of ephemerides beyond those recently claimed, and we provide an analysis of the sensitivity of our data relative to various configurations of periodic transit events.

\section{Recent Claims in the Literature to be Examined in this Work}\label{sec:previous}

The claim of an RV-detected planet in the habitable zone of Proxima by A2016 led to rapid and increased attention to this system, especially in the hopes of identifying possible transits associated with Proxima~b. In this section and in Table \ref{tbl:litephemerides}, we summarize the original and re-analyzed RV claims and associated predicted ephemerides and the claims of tentative transit-like event detections. Then, using our own extensive light curve observations presented in Section~\ref{sec:observations}, we analyze the reported detections in Section~\ref{sec:results}.  We emphasize that the photometry-based claims described below are low significance tentative detections, especially considering the apparent intrinsic variability of Proxima, and are not generally claimed by the reporting authors as confirmation of a planet transiting Proxima.

\begin{table*}[htb!]
\begin{center}
\caption{Literature Ephemerides Investigated 
\label{tbl:litephemerides}}
{\setlength{\tabcolsep}{0.30em}
\begin{tabular}{lcccccl}
\tableline
\multicolumn{1}{l}{Reference} & \multicolumn{1}{c}{Data Source} & \multicolumn{1}{c}{Period} & \multicolumn{1}{c}{$\mathrm{T}_0-2450000$} & \multicolumn{1}{c}{Depth} & \multicolumn{1}{c}{Duration}\\ 
\multicolumn{1}{l}{} & \multicolumn{1}{c}{} & \multicolumn{1}{c}{(d)} & \multicolumn{1}{c}{($\rm{BJD_{TDB}}$)} & \multicolumn{1}{c}{(\%)} & \multicolumn{1}{c}{(min)}\\
\vspace{-0.1in}\\
\tableline
\vspace{-0.1in}\\
A2016	& RV data \& analysis & 11.186$^{+0.001}_{-0.002}$ & 1634.73  &  est. 0.5 & - \\
K2017 &	RV ephemeris re-analysis &	$11.1856\pm0.0013$ &	$6678.78\pm0.56$ & est. 0.48$^{+0.14}_{-0.11}$ & [76.4]$^1$ \\
&	MOST data, Model $\mathcal{M}_1$, Signal S &	11.18467$^{+0.002}_{-0.00039}$ &	 6983.1663$^{+0.00648}_{-0.0329}$  & [1.06] & [64.8] \\
&	MOST data, Model $\mathcal{M}_2$, Signal C &	11.18725$^{+0.00012}_{-0.00016}$ &	 6980.0573$^{+0.00156}_{-0.00344}$  & [0.84] &  [64.1] \\
D2017 &	RV re-analysis &	$11.1855^{+0.0007}_{-0.0006}$ &	7383.71$^{+0.24}_{-0.21}$	& - & - \\
L2017 & MOST Signal C + BSST light curve & 11.18858 & 6801.0439 & $0.48\pm0.09$ & $82.6\pm5.3$ \\
Li2017 & Las Campanas 30 cm Robot light curve & $2-4$ & 7626.563554$^{0.001582}_{0.002355}$	& 0.46  &  $\sim60$\\
\vspace{-0.1in}\\
\tableline
\vspace{-0.1in}\\
\end{tabular}}
\tablecomments{
Quantities in square brackets were calculated from model and stellar parameters provided in K2017. All $\mathrm{T}_0$ values have been converted to \bjdtdb\ \citep{Eastman:2010} and all transit depths have been converted to percent. $^1$To determine the transit duration listed for the K2017 RV-based ephemeris, we assumed an impact parameter, $b=0$, and eccentricity, $e=0$, and stellar properties $\Mstar=0.123$\,\MS, $\Rstar= 0.145$\,\RS\ \citep{Demory:2009} and nominal planetary radius $\RP =1.06$\,\RE\ from K2017 Section 3.2.
}
\end{center}
\end{table*}

\subsection{A2016}

A2016 report the discovery of Proxima~b based on a total of 216 RV observations collected over 16 years. A subset of 54 RV observations were concentrated in a $\sim$75~day period in 2016. The RV data showed a periodic signal with reference epoch $T_0=2451634.73146$ $\mathrm{JD_{UTC}}$, a Doppler semi-amplitude $\mathrm{K}\sim1.38\pm0.21$\,m\,s$^{-1}$, eccentricity $\mathrm{e}<0.35$, and period $\mathrm{P}=11.186^{+0.001}_{-0.002}$ days, which is stable over $\sim$16 years. The corresponding minimum planet mass is $\sim$1.27~M$_{\Earth}$, and the probability of a transit orientated orbit is 1.5\%. Based on the minimum mass and an Earth-like density, the predicted transit depth is $\sim0.5\%$. Evidence is also found for an additional periodicity in the RV data in the range of $60-500$ days.

\subsection{D2017}\label{sec:Damasso2017}
D2017 undertook a re-analysis of the RV data reported by A2016 using a Gaussian process (GP) framework to mitigate the stellar correlated noise in the RV time-series. The analysis resulted in a revised orbital period $\mathrm{P}=11.1855^{+0.0007}_{-0.0006}$\,days, reference epoch $T_0= 2457383.71^{+0.24}_{-0.21}$ $\mathrm{JD_{UTC}}$, eccentricity $\mathrm{e}=0.17^{+0.21}_{-0.12}$, and Doppler semi-amplitude $\mathrm{K}\sim1.48^{+0.13}_{-0.12}$\,m\,s$^{-1}$. Their analysis dismisses the possibility of an additional planet signal in the A2016 RV data.

\subsection{K2017}
K2017 present broadband optical photometric observations of Proxima obtained with the MOST space telescope made over 12.5~days in 2014 ($\sim$2600 time-series observations) and 31 days in 2015 ($\sim$13000 observations). K2017 also re-analyzed the A2016 RV data and extracted a new RV ephemeris with $\mathrm{T}_0 = 2456678.78\pm0.56$ HJD and orbital period $\mathrm{P}=11.1856\pm0.0013$~d.

K2017 use a GP+transit model with an uninformative prior on transit phase (model $\mathcal{M}_1$) that yields four transit epochs within the MOST time-series, although one of these occurs during a data gap. This detection is referred to as signal S after phase-folding. The ephemeris derived from the events has $\mathrm{T}_0 = 2456983.1656^{+0.0064}_{-0.0330}$~HJD, which is more than 4$\sigma$ from the RV ephemeris prediction, and $\mathrm{P}=11.18467^{+0.00200}_{-0.00039}$ days. K2017 state that the observed event mid-points are ``difficult to reconcile with the radial velocity solution.''

Using a GP+transit model with an informative prior on transit phase (model $\mathcal{M}_2$) yields two events (at $\sim$2456801.06 and $\sim$2457159.05 HJD). This detection is referred to as signal C after phase-folding. The ephemeris derived from signal C has $\mathrm{T}_0 = 2456980.0554^{+0.0027}_{-0.0023}$~HJD, which is 1.5$\sigma$ consistent with the RV ephemeris, and $\mathrm{P}=11.18725^{+0.00012}_{-0.00016}$\,d. A similar model which also includes an informative prior on the radius of the planet (model $\mathcal{M}_3$) finds the same signal. K2017 conclude that HATSouth data moderately disfavor the existence of signal C at the $1-2\sigma$ level.

Finally, $\mathcal{M}_2$ was run a total of 100 times while iteratively translating the prior on T$_0$ by 0.01P, which effectively searches the full phase of the period for transit signals. This search yields three new events that are referred to as signal T after phase folding. Over 95\% of the posterior trials correspond to a grazing geometry, which is also evident from the V-shaped morphology of signal, which favors a large planet that is highly incompatible with the \texttt{Forecaster} prediction. K2017 assert that signal T would not be considered a detection even if its phase had been compatible with the RV ephemeris, so we do not consider it further in this work.

To allow for comparison with other claimed transit-like detections, we calculate transit depth and duration for signals S and C from the model $\mathcal{M}_1$ and $\mathcal{M}_2$ parameters, respectively, and include the results in Table \ref{tbl:litephemerides}.

\subsection{L2017}

L2017 report photometric observations from the Bright Star Survey Telescope (BSST) located at the Chinese Antarctic Zhongshan Station. Ten nights of observations were obtained from 29 August to 21 September, 2016.

They detect with $2.5\sigma$ confidence a transit-like event with $T_C = 2457640.1990\pm0.0017$~HJD, which is $\sim 1\sigma$ from the K2017 and $\sim 2\sigma$ from the D2017 RV predicted ephemerides. This event occurs 138~min later than predicted by the K2017 model $\mathcal{M}_2$ ephemeris.

Fitting a linear ephemeris to the two tentative K2017 signal C events and the L2017 event yields a new ephemeris with a period of $P = 11.18858$~days and $T_0 = 2456801.0439$~HJD (adopting the first $\mathrm{T_{lin}}$ value in their Table 2). The resulting transit timing variations (TTVs) relative to the linear ephemeris are in the range of $17-39$~minutes. L2017 calculate that an Earth-mass planet orbiting near a 2:1 or 3:2 mean motion resonance with Proxima~b is able to produce TTVs $\gtrsim30$ minutes
while keeping Proxima's RV$<3$\,m\,s$^{-1}$.

\subsection{Li2017}
Li2017 report one potential transit with a depth of $\sim0.5\%$ in photometric observations made over 23 nights with a robotic 30-cm telescope at Las Campanas Observatory. The modeled mid-transit time is $\mathrm{T}_C= 2457626.5635537^{+0.0015813}_{-0.0023548}$\,\bjdtdb\ and the duration is about one hour. Li2017 show that if the event is indeed caused by a planet transiting Proxima, the transit model prefers a $2-4$~day orbit. Furthermore, the planet mass would need to be $<0.4$\,\ME\ to avoid detection by the A2016 RVs.

\section{New Observations and Reductions}\label{sec:observations}

\subsection{Observations}

We have conducted an extensive photometric monitoring campaign of Proxima using multiple ground-based observatories from 2006 to 2017. The campaign conducted observations routinely each year, except for a gap in observations from 2009 to 2012. In total, we obtained 329 nights of time-series photometric observations that resulted in light curves of at least 1.5 hours in duration (most are 3--8 hours in duration) after data processing and cleaning (see Section \ref{sec:datareduction}). In this section, we describe the observations, which are summarized in Table~\ref{tbl:photobs}.  To the best of our knowledge, this is the longest duration transit study of Proxima to date. 

In the present work, we restrict our analyses to a subset of 96 observations that coincide with the predicted times of transit from previously published claims (see Section~\ref{sec:previous} and Table~\ref{tbl:litephemerides}). The 96 light curves are presented in the Appendix\ref{sec:appendix} and will be provided along with the full set of light curves in machine readable format in Paper~II.

\begin{table*}[htb!]
\begin{center}
\caption{Summary of Photometric Observations Analyzed in this Work and in Paper~II\label{tbl:photobs}}
{\setlength{\tabcolsep}{0.30em}
\begin{tabular}{lcccllccr}
\tableline
\multicolumn{1}{l}{Telescope Name} & \multicolumn{1}{c}{Aperture} & \multicolumn{1}{c}{FOV} & \multicolumn{1}{c}{Plate-Scale} & \multicolumn{1}{c}{Start Date} & \multicolumn{1}{c}{End Date} & \multicolumn{1}{c}{Exp. Time} & \multicolumn{1}{c}{Filter} & \multicolumn{1}{r}{\# Obs} \\ 
\multicolumn{1}{l}{} & \multicolumn{1}{c}{(m)} & \multicolumn{1}{c}{(arcmin$^2$)} & \multicolumn{1}{c}{(arcsec pixel$^{-1}$)} & \multicolumn{1}{c}{(UT)} & \multicolumn{1}{c}{(UT)}  & \multicolumn{1}{c}{(sec)} & \multicolumn{1}{c}{ } & \multicolumn{1}{c}{(nights)}\\
\vspace{-0.1in}\\
\tableline
\vspace{-0.1in}\\
RAE  & 0.35  & 10.4$\times$10.4 & 1.2  &  2006 May 24 &  2008 Feb 25 &  20  & R & 23\\
RCOP    & 0.4  & 24.2$\times$16.3 & 0.76  &  2014 Feb 13 &  2014 Aug 23 &  16-20  & R & 30\\
Prompt 1        & 0.4  &  9.64$\times$9.64 & 0.9  & 2013 Aug 17 &  2015 Apr 22 &  16-20  & R & 40\\ 
Prompt 2    & 0.4  & 21$\times$14 & 0.41  &  2013 Aug 21 &  2017 Mar 07 &  15--20, 65  & R,G & 50\\
Prompt 4       & 0.4  & 10$\times$10 & 0.59  &  2014 Mar 07 &  2015 May 11 &  15-20  & R & 50\\
Prompt 5       & 0.4  & 10.25$\times$10.25 & 0.59  &  2014 Mar 16  &  2016 Mar 29 &  18-20  & R & 10\\
Prompt 8    & 0.6  & 22.6$\times$22.6 & 0.69  &  2014 Jun 20 &  2015 Mar 15 &  16-18  & R & 3\\
Prompt SS01       & 0.42  & 15.6$\times$15.6 & 0.9  & 2014 Feb 23 &  2015 May 05 &  15-20  & R & 46\\
Prompt SS02       & 0.42  & 15.6$\times$15.6 & 0.9  & 2014 Feb 23 &  2014 Jul 30 &  17-20  & R & 18\\
Prompt SS03       & 0.42  & 15.6$\times$15.6 & 0.9  & 2014 May 08 &  2014 Aug 14 &  15-20  & R & 40\\
Prompt SS04       & 0.42  & 15.6$\times$15.6 & 0.9  & 2013 Sep 02 &  2013 Sep 13 &  20  & R & 2\\
Hazelwood      & 0.32  & 18$\times$12 & 0.73  & 2017 Mar 18 &  2017 Jun 16 &  5-12  & Ic & 6\\
Ellinbank       & 0.32  & 20.2$\times$13.5 & 1.12  & 2017 Jun 16 &  2017 Jul 30  &  14-18  & R & 5\\
Mt. Kent CDK700       & 0.7  & 27.3$\times$27.3 & 0.40  & 2017 Jun 20 &  2017 Jul 25 &  20-25  & I & 3\\
ICO        & 0.235  &  16.6$\times$12.3 & 0.62  & 2017 Mar 18  &       2017 May 14  &  15-30  & I & 3\\

\tableline
\end{tabular}}
\end{center}
\end{table*}

\subsubsection{Observing Strategy}\label{sec:observingstrategy}

Because our survey started in 2006, almost all of our observations were conducted before the A2016 RV-discovered planet was announced. Thus, we were generally conducting a blind search for transits of Proxima through 2016. After the RV ephemeris was announced, we targeted the D2017 predicted times of transit, allowing for about a day of uncertainty in the transit window. 

Proxima is known to be a flare star \citep{Shapley:1951, Walker:1981}, and M-dwarf flares are known to be much brighter in blue passbands compared to red passbands \citep{Kowalski:2016}. Although Proxima's flares are prominent across the UV and optical bands, they are indeed stronger in the blue compared to the underlying stellar photosphere \citep{Walker:1981}, so we targeted the $R$ passband and redder to minimize the impact on our photometric observations. Even so, contamination from systematics, flares, and low energy flares that are now predicted to occur about every $\sim20$ minutes at the 0.5\% level \citep{Davenport:2016}, is significant in our photometric data. Thus we needed a large number of observations to help improve the sensitivity of our data to periodic transit signals predicted to have a depth similar to the amplitude of variations common in our data.

\subsubsection{Perth Observations}

Our 2006 to 2008 observations were obtained using the Real Astronomy Experience (RAE) Robotic Telescope at the Perth Observatory in Bickley, Western Australia, which was primarily used for astronomy education in the ``Hands on Universe" program \citep{Fadavi:2006}\footnote{www.hou.org}. The telescope was a Schmidt-Cassegrain design with an aperture of 0.35\,m and was equipped with an Apogee Ap7 CCD camera and $BVRI$ filters. Despite having a plate scale of $1.2 \arcsec$ pixel$^{-1}$, the seeing was generally at least $3\arcsec$, so under-sampling was not typically a problem. The RAE telescope was not guided, so periodic re-pointing of the telescope by the robot was required to keep the field centered on the detector. Additional telescope specifications are provided in Table~\ref{tbl:photobs}.

The RAE observations were conducted in an $R$ filter with exposure times of 20\,s. The CCD readout time was 11\,s, yielding an effective cadence of 31\,s per exposure. Proxima was the brightest star in the RAE field and the ADU counts varied between 22,000 and 35,000, which ensured that the frames were not saturated but still well exposed. Bias, dark, and flat-field frames were applied automatically to each science exposure by the telescope system.  The telescope was operated remotely through the internet. 

\subsubsection{Skynet Observations}
 
The great bulk of our observations were obtained using the Skynet world-wide network of remotely operated 0.4~m and 0.6~m telescopes \citep{Reichart:2005}. Our observations were obtained using telescopes located at Cerro Tololo, Chile (Prompt 1, 2, 4, 5, and 8), Siding Springs, New South Wales, Australia (Prompt SS01, SS02, SS03, and SS04) and Perth, Western Australia (referred to as RCOP, hereafter).  All of our observations were obtained with the 0.4~m telescopes, except for three observation runs with the 0.6~m Prompt 8 telescope. In general, the Skynet telescopes were not guided, so periodic re-pointing of the telescope by the robot was required to keep the field centered on the detector. Additional Skynet telescope specifications are included in Table~\ref{tbl:photobs}.

Frames were exposed in an $R$ filter with integration times ranging from 15 to 20~s, except for one 2017 Prompt 2 observation with an integration time of 65~s, which used a generic green filter (listed as G in Table ~\ref{tbl:photobs}). We adopted the standard Skynet calibrated data, which includes dark, bias, and flat-field corrections.

\subsubsection{KELT-FUN Observations}

We collected observations from the Kilodegree Extremely Little Telescope (KELT; \citealt{Pepper:2007,Pepper:2012}) Follow-Up Network (KELT-FUN; \citealt{Collins:2018}) based on the A2016 RV-based ephemeris. We used the {\tt Tapir} software package \citep{Jensen:2013} to schedule the KELT-FUN observations. KELT-FUN members contributed a total of 18 light curves from March, 2017 to July, 2017. The KELT-FUN observations are identified in Table~\ref{tbl:photobs} as Hazelwood, Ellinbank, Mt. Kent CDK700, and Ivan Curtis Observatory (ICO). The single Prompt~2 G-band observation in March of 2017 was also contributed by a KELT-FUN member using Skynet time allocated to the KELT project. KELT-FUN telescope specifications are included in Table~\ref{tbl:photobs}. Image calibration included dark, bias, and flat-field corrections.
\\
\\
\subsection{Data Reduction\label{sec:datareduction}}

To achieve the photometric precision needed to detect a $\sim0.5\%$ transit-like event in our ground-based observations, we require differential photometry to compensate for the adverse effects of the atmosphere. However, it is difficult to directly compare differential photometry across multiple nights and multiple telescopes due to telescope pointing inaccuracies combined with imperfect flat-field compensation, long term changes in comparison star brightness, differences in the comparison stars available on the detector, chromatic differences in atmospheric transparency, differences in atmospheric scintillation, changes in telescope focus, etc. 
In our case, telescope guiding was not implemented for the RAE and Skynet observations, so the significant changes in the placement of the field on the detector throughout a time-series limited the number of comparison stars available on the detector for the entire sequence. Fewer comparison stars generally results in lower photometric precision and higher levels of systematics. The comparison star ensemble problem would typically be compounded across multi-night differential photometry, if trying to use the same ensemble to directly compare the differential light curves, so we allowed for different comparison star ensembles for each night of observations.

To overcome the different calibration of the multi-night differential photometry, we chose to process and then normalize each light curve separately, such that the final mean value is 1.0. We discuss the data processing below. While normalizing the observations allows a direct comparison of multi-night light curves, we acknowledge that a real event could be obscured if the duration is longer than $\sim 50\%$ of the duration of the light curve. To minimize this issue, we generally required light curves to be at least $\sim2.5$ hours long before data processing. In some cases, data processing reduced the light curve duration, so we set a hard lower limit of 1.5 hours of coverage and dropped light curves with a shorter final duration.

AstroImageJ (AIJ; \citealt{Collins:2017}) was used to perform differential photometry on all data sets. The images for each night were inspected manually and frames contaminated with aircraft, satellite, clouds, etc, that might cause photometric inaccuracies, were discarded. In general, we find that selecting comparison stars that have $\pm50\%$ of the brightness of the target star produces the least systematics in the target star light curve. In addition, we generally find that balancing the number of ensemble integrated counts from comparison stars fainter and brighter than the target star reduces light curve systematics even more. However, Proxima is the brightest star in the field of our detectors in our filter bands, so only fainter comparison stars were available for the ensemble. For this work, we first selected all comparison stars that are at least 50\% as bright as Proxima. However, in many cases (in particular for the RAE and Skynet telescopes with relatively small fields of view), only one or two comparison stars with the desired brightness were available on the detector throughout the time series, so we generally selected the $\sim5$ brightest stars available for the ensemble, avoiding stars that showed significant variability. For most image sequences, we used an aperture with an $8$ pixel radius, but the radius varied depending on differences in detector pixel scales, seeing, and telescope focus. 

To minimize the effects of chromatic differential airmass trend and long term stellar variability, we assumed a flat light curve model and performed a linear detrend using, at a minimum, airmass and time. In some light curves with a strong correlation between the x- and/or y-centroid of the target star location on the detector and variability in the light curve, we performed a linear detrend using the x- and/or y-centroid locations of the target star. In some cases we detrended using sky background, full-width half-maximum of the stellar point spread function, and/or the total number of comparison star net integrated counts. If a telescope meridian flip or tracking jump resulted in an correlated change in the photometric baseline, we fitted and realigned the baseline at that point. This method of detrending will help to minimize false event detections due to potentially large step functions at the ends of individual light curves when we perform our periodic transit search for Paper~II. The normalizing and detrending process also minimizes the effects of Proxima's $82.6\pm0.1$ day rotation period \citep{CollinsJM:2017} and seven year stellar cycle \citep{Wargelin:2017}. While we acknowledge that detrending using this method may reduce or enhance the significance of transit-like events, we have visually compared each undetrended light curve with its detrended version, and can confirm that the adverse affects are minimal. In Paper II or a followup paper, we also intend to investigate de-weighting data near the edges of individual light curves to potentially reduce the need for detrending and/or to improve the periodic transit search results. 

The light curves have between 79 and 1468 data points each, with an average of 483 data points. Assuming perfectly Gaussian distributed data sets containing 79, 483, and 1468 data points, Chauvenet's criterion \citep{Chauvenet:1960} specifies that values beyond 2.7, 3.3, and 3.7$\sigma$ from the mean, respectively, should be considered outliers. Therefore, to remove large flares and other photometric outlier data points, we elected to perform a uniform iterative $3\sigma$ cut on each individual light curve using AIJ. After each $>\pm3\sigma$ outlier point was removed, AIJ detrended and normalized the data again, and the process was repeated until no $3\sigma$ outlier data points remained. We visually inspected each light curve before and after the cuts to verify that the cleaning operation did not remove any obvious transit-like events in our data. After very strong flares, we also removed additional data points in the light curve that were not removed by the $3\sigma$ cut, but that were obviously affected by the rising or decaying flare signal. We also removed short segments of data that were separated in time from the main cluster of data, and that likely did not share the same baseline differential flux value due to a telescope meridian flip or a large instantaneous shift of the field on the detector.

After the data were cleaned, 329 light curves (167,445 photometric data points) having a duration of 1.5 hours or longer remained in our sample. Even after detrending and $3\sigma$ cleaning of the light curves, some had very large oscillatory or other variations that were not consistent with a transit signal, or would have prevented detection of an underlying $\sim0.5-1.0\%$ transit-like event. We sought to exclude these light curves using a statistical cut. 

\begin{figure}[!htb]
\centering \includegraphics[width=\columnwidth, trim=0cm 0.0cm 0cm 0.0cm, clip=true]{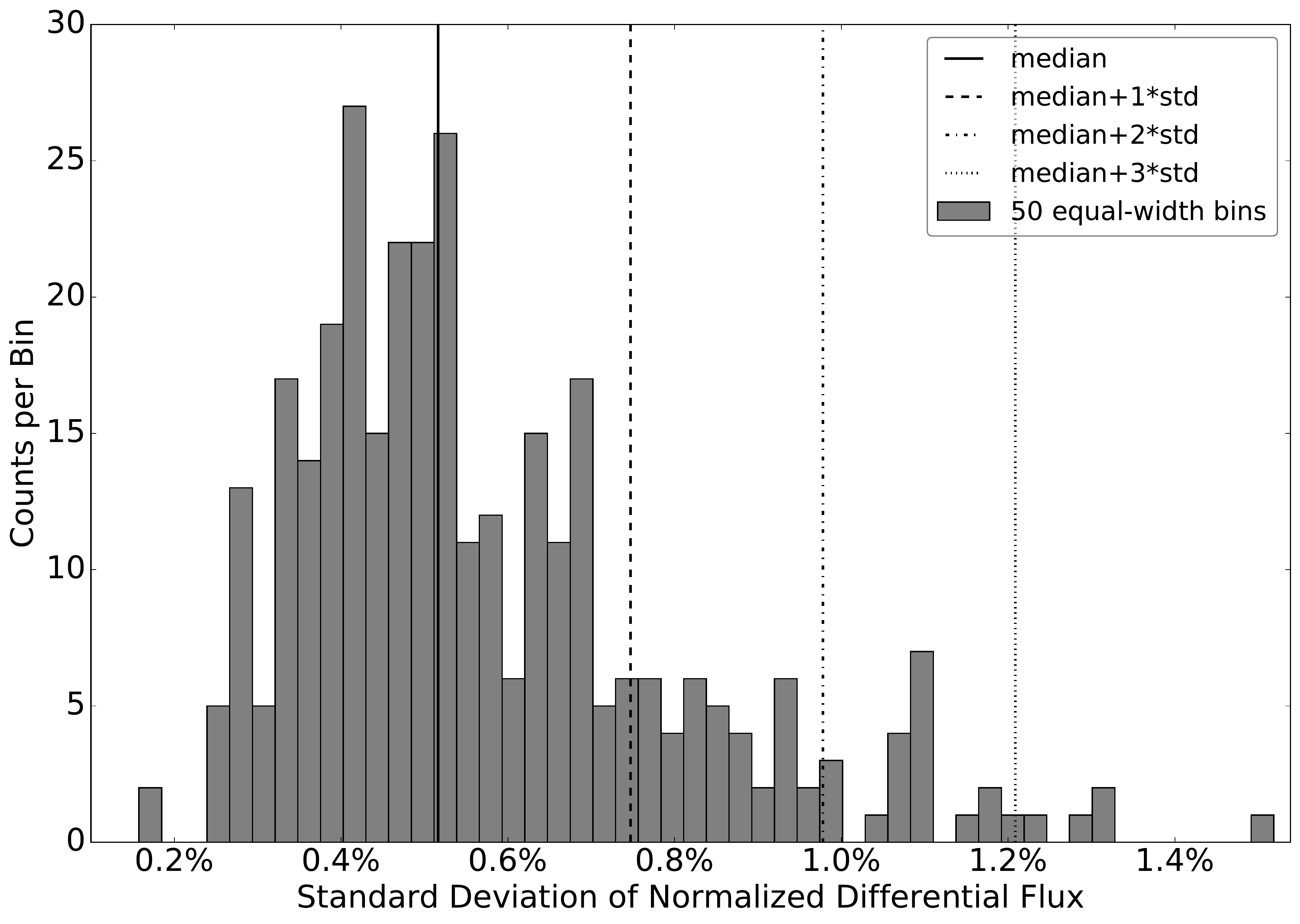}
\caption{Histogram of the standard deviations of our 329 individual light curves. The solid line marks the median of the distribution at 0.516\%. The distribution has a standard deviation of 0.230\%. The long-, medium-, and short-dashed lines mark the values of median plus 1, 2, and 3 times the standard deviation, respectively. }
\label{fig:rms}
\end{figure}

\begin{figure}
    \centering
    \includegraphics[width=\columnwidth, trim=1.0cm 0.0cm 1.0cm 0.0cm, clip=true]{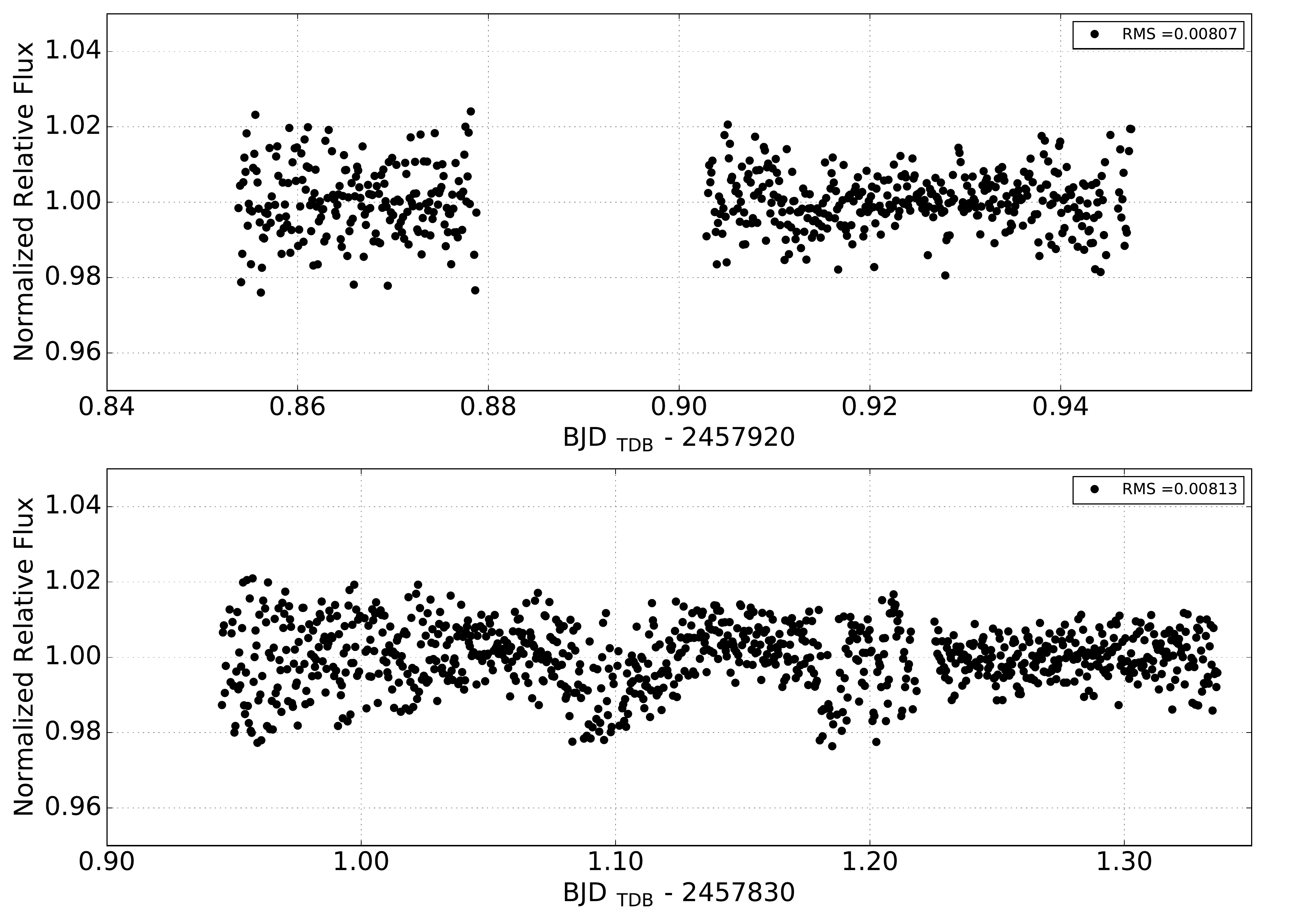}
    \caption{Examples light curves retained in our data set, despite having standard deviation above our threshold. (Top Panel) The KELT-FUN Hazelwood Observatory light curve from UT 2017 June 16 is relatively flat but has scatter above our threshold. (Bottom Panel) The KELT-FUN ICO light curve from UT 2017 March 18 has a transit-like feature that contributes to the high scatter.}
    \label{fig:high_scatter_examples}
\end{figure}

Figure \ref{fig:rms} shows a histogram of the 329 standard deviations of the individual cleaned and detrended light curves. The standard deviations range from $\sim0.17-1.5\%$ and have a median value of 0.516\%. The distribution has a standard deviation of 0.230\%. The light curves with standard deviation above the distribution's median value plus the distribution's standard deviation (0.746\%) were flagged to be removed from the analysis. We examined each of the flagged light curves and found that several had large standard deviation due to either a transit-like event (although generally deeper than predicted for Proxima b) or a relatively flat light curve with white-noise-like scatter above the threshold. We retained those two types of light curves in our sample, despite the large standard deviation. Two examples of light curves retained in our sample, despite being above our standard deviation threshold are shown in Figure \ref{fig:high_scatter_examples}. The final light curve count in our study sample is 262, which includes a total of 127,733 photometric data points.

\section{Results}\label{sec:results}

From the 262 light curves in the final sample, the Appendix\ref{sec:appendix} describes and displays the subset of 96 light curves that coincide with the published ephemerides described in Section~\ref{sec:previous} and Table~\ref{tbl:litephemerides}. The full set of 262 light curves and their analysis will be presented in Paper~II.

\subsection{Light Curves in Relation to RV-based Ephemerides}

In total, there are 85 light curves from the final sample that contribute data within $2\sigma$ of the K2017 RV-based ephemeris. The light curves are phased to the K2017 RV-based ephemeris and displayed in the  Appendix\ref{sec:appendix}, along with the K2017 MOST light curves and the L2017 BSST light curves.

\begin{figure*}[!htb]
\centering \includegraphics[width=\linewidth, trim=0.0cm 0.0cm 0.0cm 0.0cm, clip=true]{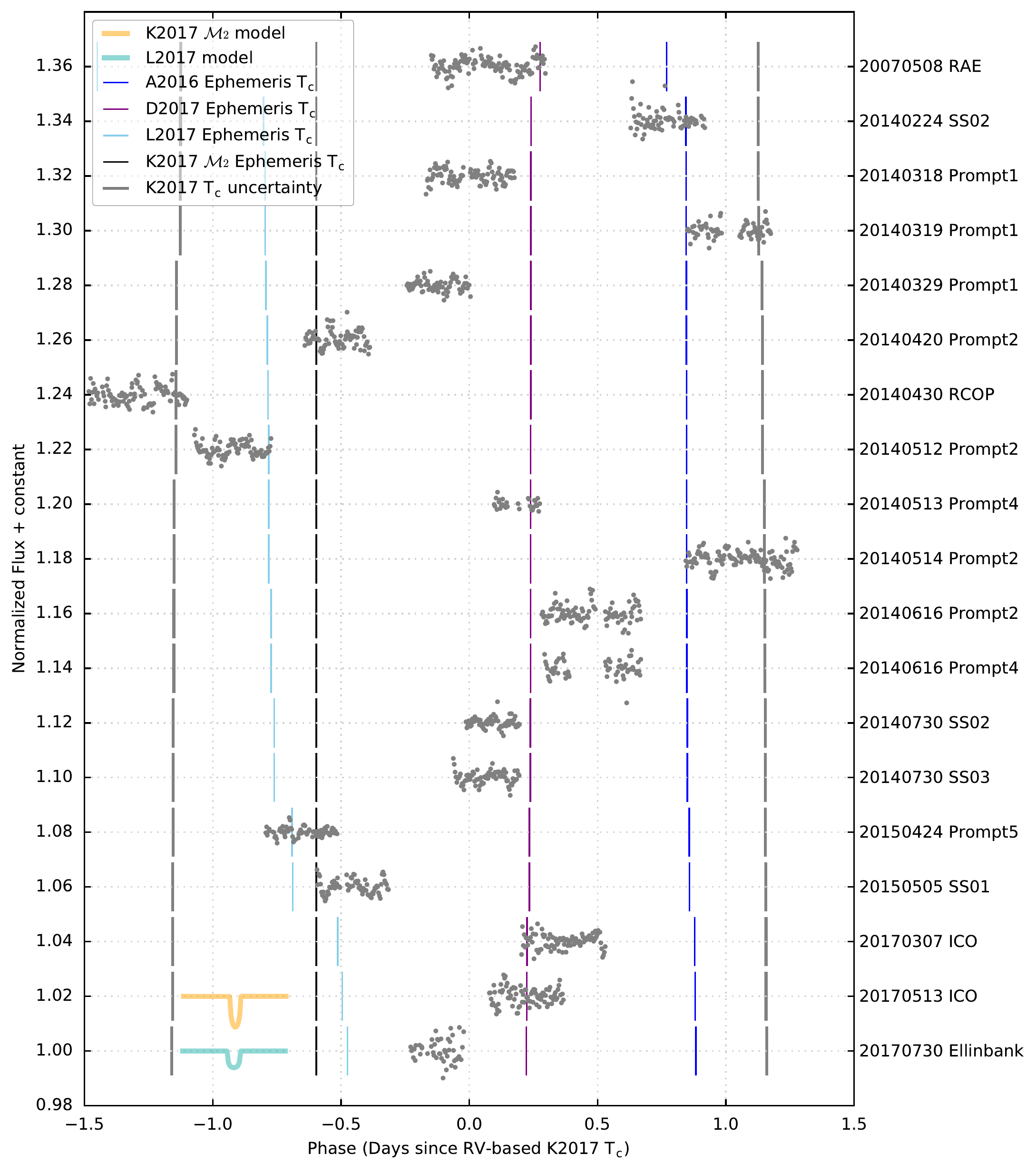}
\caption{A subset of 19 light curves from this work phased to the K2017 RV-based ephemeris. The nominal transit center times predicted by the K2017 RV-based ephemeris are located in the center of the figure at phase zero. The grey vertical solid lines mark the $\pm2\sigma$ uncertainty in the ephemeris. The blue, magenta, light blue and black vertical solid lines mark the transit center times predicted by the ephemerides of A2016, D2017, L2017, and the K2017 model $\mathcal{M}_2$ (Signal C), respectively. The K2017 $\mathcal{M}_2$ and L2017 models are shown to scale in the lower left corner as solid orange and light green solid lines, respectively.} All 85 light curves contributing data within $\pm2\sigma$ of the K2017 RV-based ephemeris are presented in the  Appendix\ref{sec:appendix}.
\label{fig:snippet}
\end{figure*}

The vertical scale of the light curves in the Appendix\ref{sec:appendix} is compressed to accommodate the large number of light curves. To elucidate the level of post-detrended residual variations in the light curves, Figure \ref{fig:snippet} shows a subset of 19 light curves that fall within $2\sigma$ of the K2017 RV-based ephemeris.

Each light curve has been shifted on the vertical axis for clarity, and date of observation and telescope identification, as defined in Section \ref{sec:observations} and Table \ref{tbl:photobs}, are displayed on the right-hand vertical axis. The light curve data are binned in 5 minute intervals.

The center of Figure \ref{fig:snippet}, labeled as phase zero, corresponds to the nominal predicted transit center at each epoch of displayed data according to the K2017 RV-based ephemeris. The grey vertical bars at $\sim\pm1.2$ days span the width of the $2\sigma$ uncertainty, and varies depending on the amount of time since the reference epoch, T$_0$, due to the cumulative uncertainty in the period.

Also shown are the transit centers at each displayed epoch, extracted from the other literature ephemerides listed in Table \ref{tbl:litephemerides}, after phasing to the K2017 RV-based ephemeris. The transit centers predicted by the A2016 RV-based ephemeris are displayed as blue vertical lines. The nominal A2016 transit centers are inconsistent with the K2017 RV-based ephemeris at a level of $\sim1.5\sigma$, primarily due to an offset in the reference epoch. The transit centers predicted by the D2017 ephemeris are displayed as magenta vertical solid lines. The D2017 ephemeris is highly consistent with the K2017 RV-based ephemeris, relative to the uncertainty. The K2017 Signal C and L2017 ephemerides are shown as black and light blue vertical bars, respectively. Both sets of predicted transit centers precede the K2017 RV-based ephemeris by about $1\sigma$.

The light curves exhibit a variety of behaviors. In some cases there is no clear evidence for a transit, within the noise of the data (e.g., light curves ``20140329 Prompt1'' and ``20140730 SS02''). In other cases, there is some variability of amplitude comparable to that found by other authors, but which does not have a shape that is generally consistent with a transit, or more specifically, with the previously claimed transits (e.g., light curves ``20070508 RAE'' and ``20170307 ICO''). Finally, there are some cases in which there is variability observed that could potentially be regarded as consistent with the previously claimed transit-derived models, although the transit center phase is not consistent with the models (e.g., light curve ``20140514 Prompt2''). 

\begin{figure}
\includegraphics[width=\linewidth, trim=0.0cm 0.0cm 0.0cm 0.0cm, clip=true]{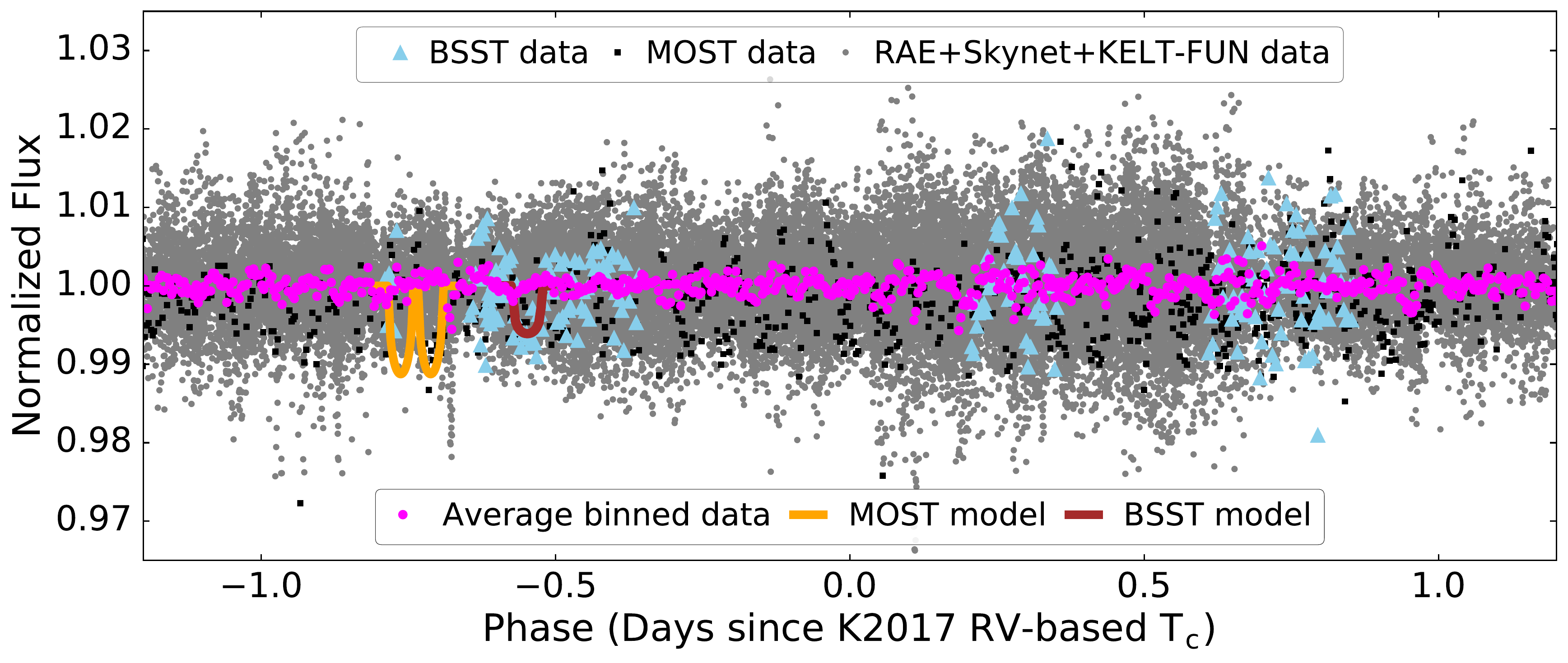}
\caption{All light curve observations, including those from the literature and those newly obtained by us, folded on the K2017 RV-based ephemeris. The data from this work are displayed as grey dots, and after combining and binning at five minute intervals, as magenta dots. The K2017 MOST data are also displayed as black squares, and the L2017 BSST data are shown as light blue triangles. The K2017 Signal C transit models are displayed as orange solid lines. The L2017 BSST transit model is displayed as a brown solid line. There are no obvious periodic transit signals, at the depth of the plotted models, evident within the noise of the binned data.}
\label{fig:kipping_rv_fold}
\end{figure}

Since the $2\sigma$ uncertainty in the RV-based transit ephemerides corresponds to a time window of approximately $\pm1.2$~days, the ground-based light curve observations presented here cannot individually span the entire time window within which transits might be expected to occur. However, after combining and phase-folding all 85 light curves from this work, the full $\pm2\sigma$ phase range has complete coverage. Figure~\ref{fig:kipping_rv_fold} shows the full phase range with the data from this work displayed as grey dots. The data are combined and binned at five minute intervals and displayed as magenta dots. The K2017 MOST data are also displayed as black squares, and the L2017 BSST data are shown as light blue triangles. The K2017 Signal C transit models are displayed as solid orange lines. The L2017 BSST transit model is displayed as a solid brown line. There are no obvious transit signals, at the depth of the plotted models evident within the noise of binned data. Note, however, that the significance of any transit signals following the other ephemerides described in Section \ref{sec:previous} would be significantly reduced due to the skewing of the transit alignments as a result of the slightly different periods compared to the K2017 RV-based ephemeris used to phase the data and transit models.

\subsection{Light Curves in Relation to K2017 Signals C and S Ephemerides}\label{sec:signalsCandS}

\begin{figure}[!htb]
\centering \includegraphics[width=\columnwidth, trim=0.0cm 0.0cm 0.0cm 0.0cm, clip=true]{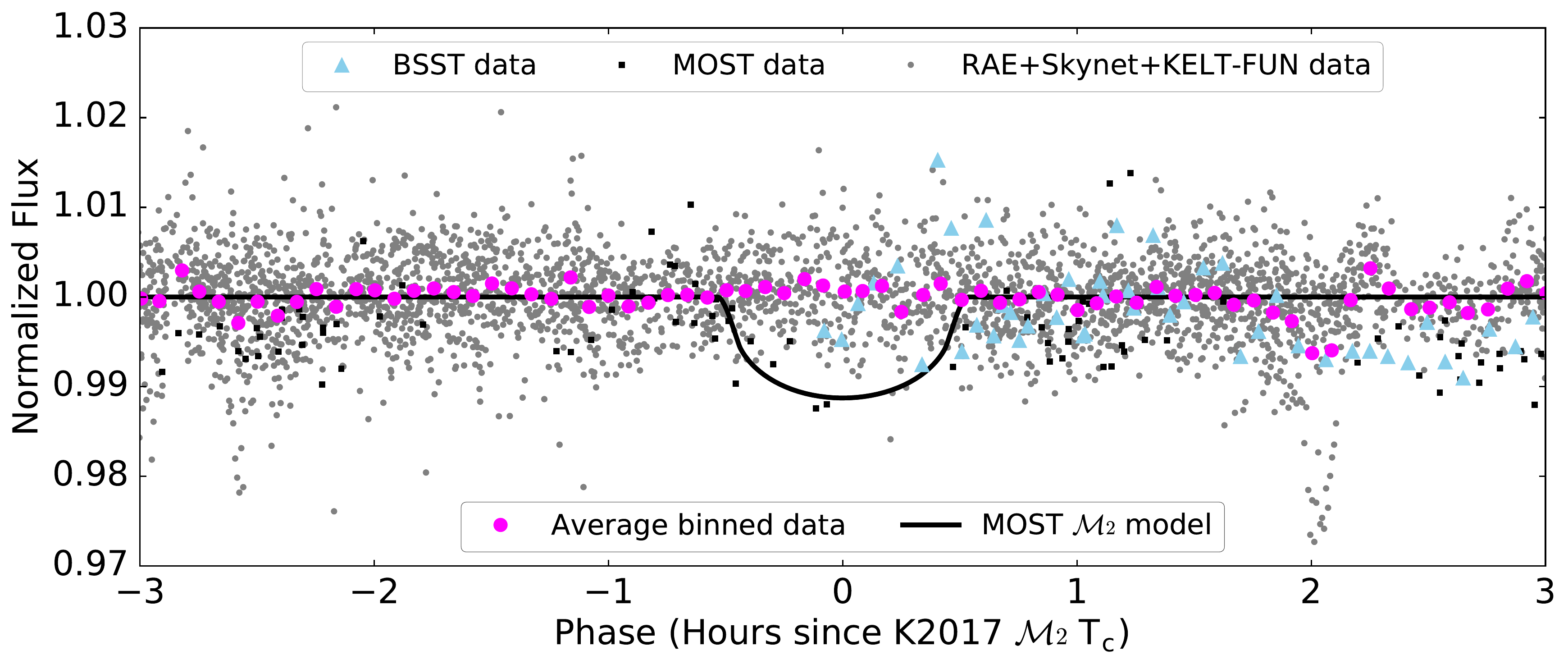}
\caption{All light curve observations, including those from the literature and those newly obtained by us, folded on the K2017 Model~$\mathcal{M}_2$ ephemeris. The phase range displayed is $\pm 3$ hours from the Model $\mathcal{M}_2$ transit center time. Light curves from this work are displayed as grey dots, and after combining and binning at five minute intervals, as magenta dots. The MOST data are shown as black squares and the BSST data are displayed as light blue triangles. The K2017 $\mathcal{M}_2$ transit model is displayed as a black solid line.}
\label{fig:kipping_M2_fold}
\end{figure}

K2017 reported a transit-like event detection, referred to as Signal C, that is within the $2\sigma$ range of their RV-based ephemeris. To check for evidence of periodic transits in our data corresponding to signal C events, our data are phase folded using the corresponding Model $\mathcal{M}_2$ ephemeris and displayed as grey dots in Figure \ref{fig:kipping_M2_fold}. The data are combined and binned at five minute intervals (after phased folding) and displayed as magenta dots. The binned data have a standard deviation of $\sim 0.20\%$, well below the 0.84\% depth of the $\mathcal{M}_2$ model (displayed as a black solid line), but there is no obvious transit-like event in our phased data. The MOST data are also displayed as black squares, and the BSST data are displayed as light blue triangles. The lack of an obvious transit-like signal in our data, relative to the depth predicted by Model $\mathcal{M}_2$ is strong evidence that Signal C was not caused by a transiting exoplanet in a periodic orbit.

\begin{figure}[!htb]
\centering \includegraphics[width=\columnwidth, trim=0.0cm 0.0cm 0.0cm 0.0cm, clip=true]{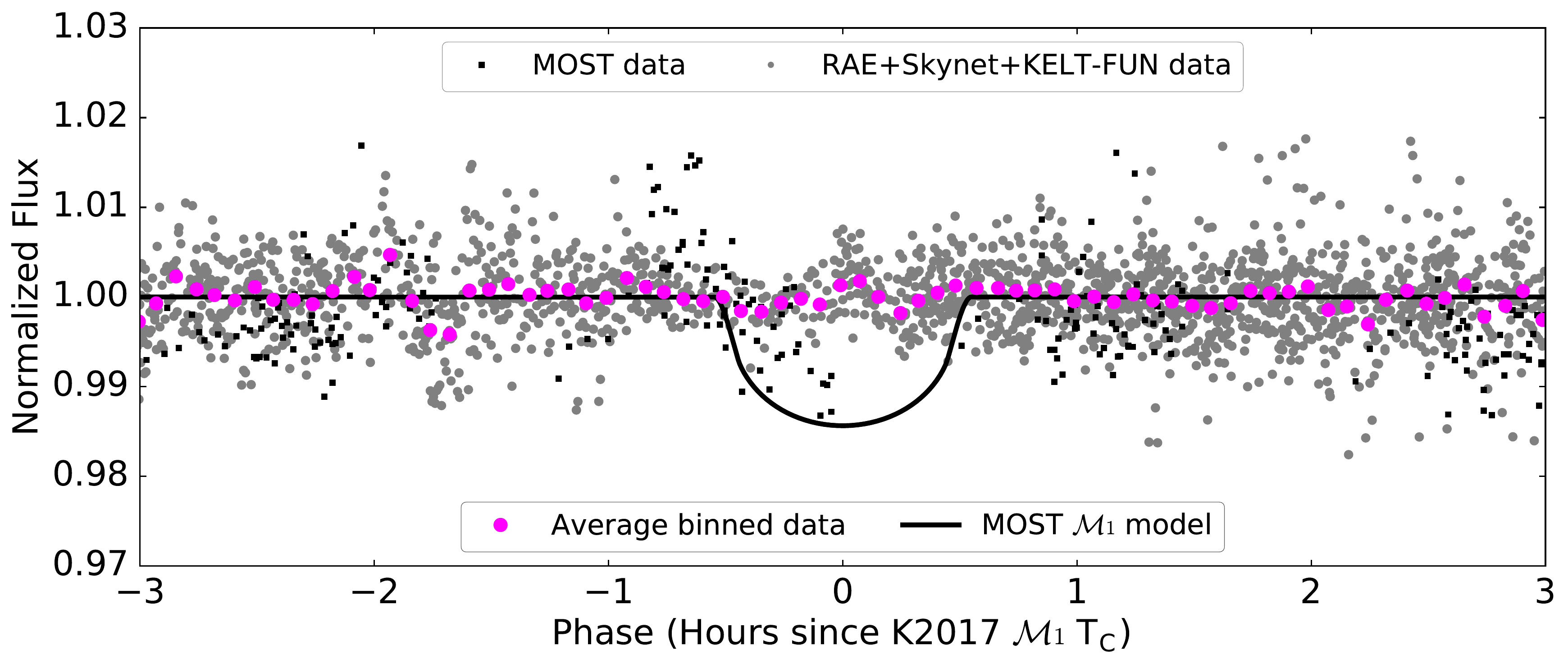}
\caption{All light curve observations, including those from the literature and those newly obtained by us, folded on the K2017 Model~$\mathcal{M}_1$ ephemeris. The phase range displayed is $\pm 3$ hours from the Model $\mathcal{M}_1$ transit center time. Light curves from this work are displayed as grey dots, and after combining and binning at five minute intervals, as magenta dots. The MOST data are shown as black squares and the $\mathcal{M}_1$ transit model is displayed as a black solid line. No BSST data contribute to the displayed phase range.}
\label{fig:kipping_M1_fold}
\end{figure}

K2017 also reported a transit-like event detection, referred to as Signal S, that is outside the $2\sigma$ range of their RV ephemeris, which they considered spurious. To check for evidence of periodic transits in our data corresponding to signal S events, the data are phase folded using the corresponding Model $\mathcal{M}_1$ ephemeris and displayed as grey dots in Figure \ref{fig:kipping_M1_fold}. The data are combined and binned at five minute intervals (after phased folding) and displayed as magenta dots. The binned data have a standard deviation of $\sim 0.21\%$, well below the 1.06\% depth of the $\mathcal{M}_1$ model (displayed as a black solid line), but there is no obvious transit-like event in our data. The normalized MOST data are also displayed as black squares, and have been shifted vertically so that they approximately align with the $\mathcal{M}_1$ light curve model near to and during the time of the event. No BSST data contribute within $\pm3$ hours of the Model $\mathcal{M}_1$ ephemeris. The lack of an obvious transit-like signal in our data supports the K2017 conclusion that Signal S is spurious. 

\subsection{Light Curves in Relation to L2017 TTV Ephemeris}

\begin{figure}[!htb]
\centering \includegraphics[width=\columnwidth, trim=0.0cm 0.0cm 0.0cm 0.0cm, clip=true]{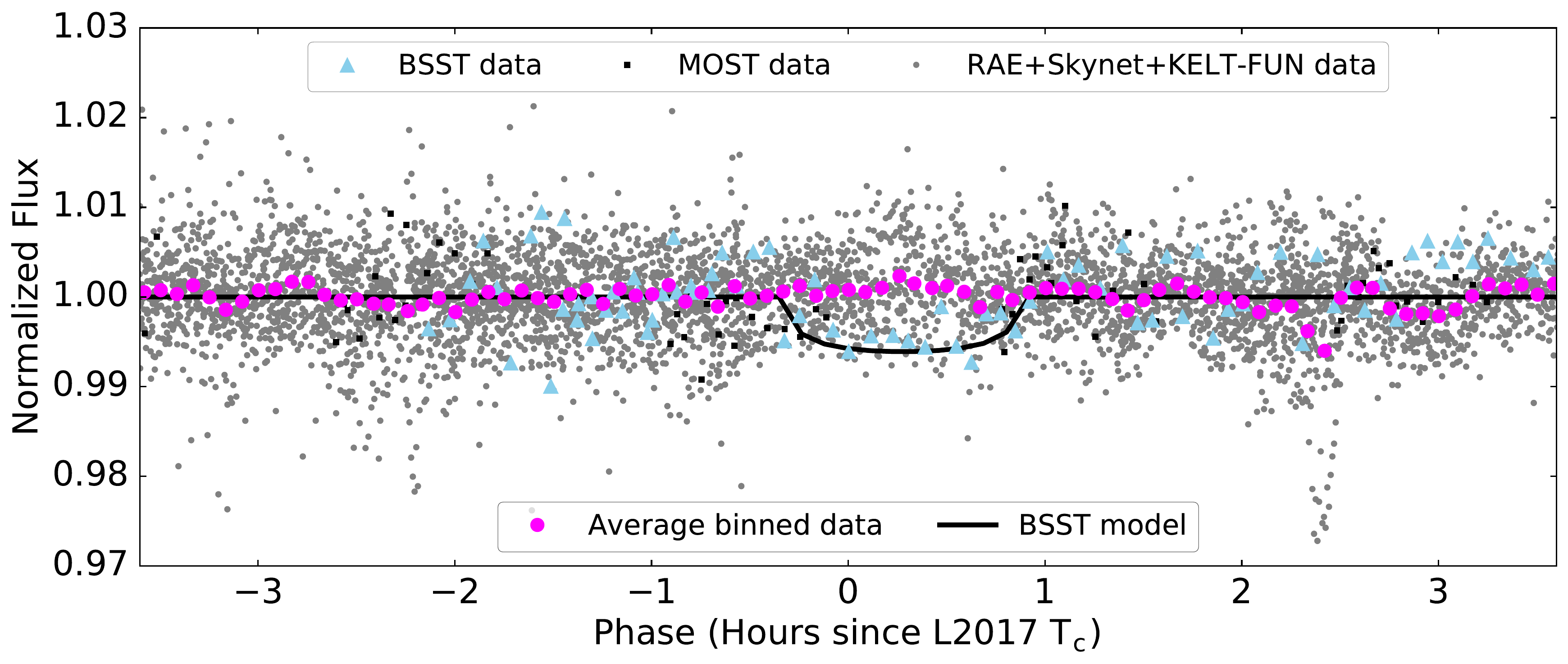}
\caption{All light curve observations, including those from the literature and those newly obtained by us, folded on the L2017 ephemeris. The phase range displayed is $\pm 3$ hours from the L2017 transit center time. The data are displayed as described for Figure~\ref{fig:kipping_M2_fold}, except that the L2017 transit model is displayed as a black solid line.}
\label{fig:liu_fold}
\end{figure}

L2017 combined a single transit-like event detection in their BSST data with the two K2017 Signal C events and found the best fit linear ephemeris, which has a slightly longer period than the Signal C ephemeris. We present the observations from this work and the MOST and BSST observations phased to the L2017 ephemeris in Figure~\ref{fig:liu_fold}. The data are displayed as described in Section~\ref{sec:signalsCandS} and Figure~\ref{fig:kipping_M2_fold}, except that the L2017 transit model is displayed as a black solid line. Our binned data have a scatter of $\sim 0.20 \%$, which is below the 0.5\% depth of the claimed transit. We see no evidence of a periodic 0.5\% deep transit signal in our binned data. In fact, there is an apparent slight brightening in our light curve during the predicted transit event, which we discuss further below.

\begin{figure}[!htb]
\centering \includegraphics[width=\columnwidth, trim=0.0cm 0.0cm 0.0cm 0.0cm, clip=true]{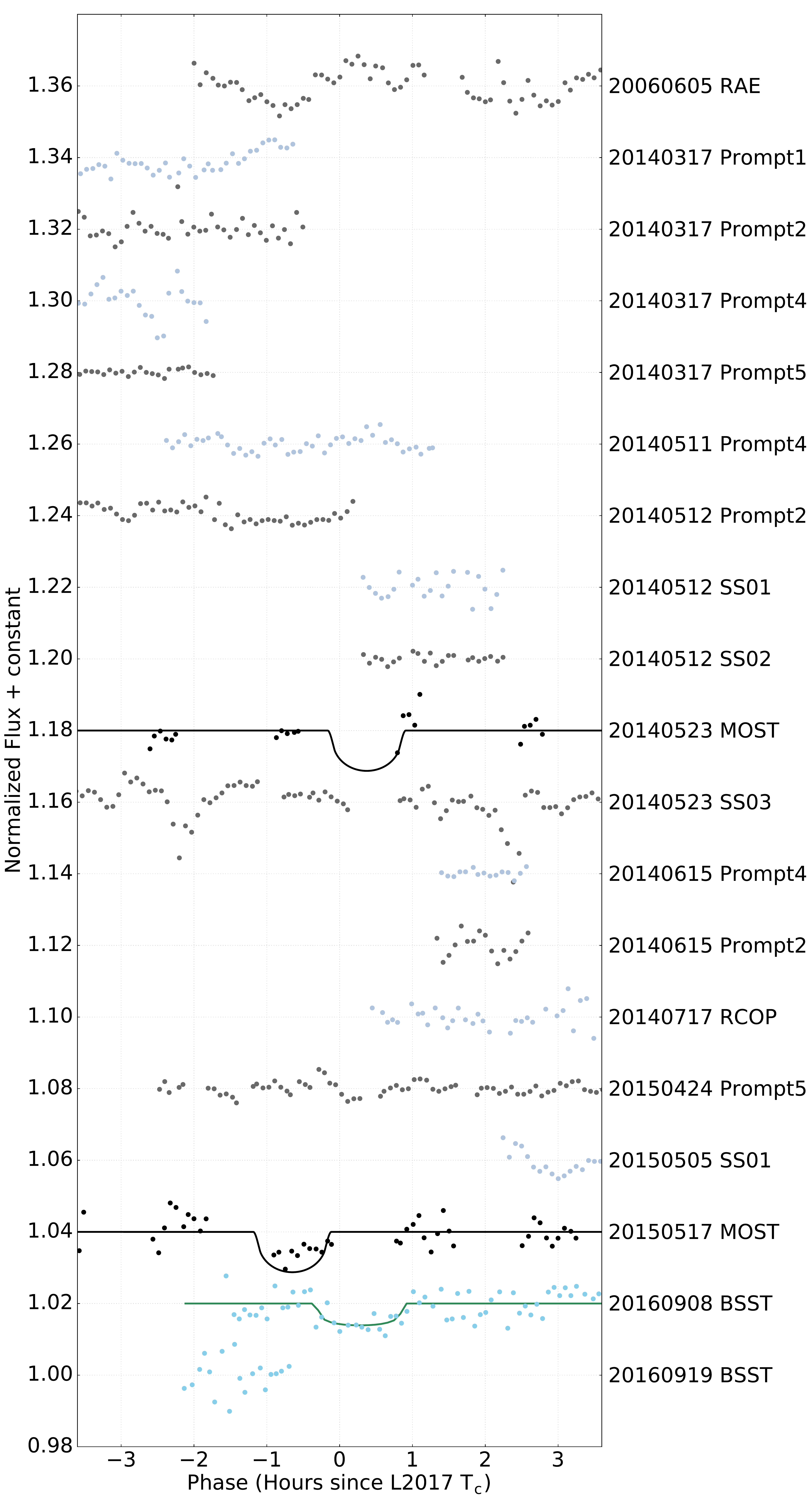}
\caption{All light curve observations, including those from the literature and those newly obtained by us, folded on the L2017 ephemeris. The phase-folded constituent light curves in Figure~\ref{fig:liu_fold} are shifted relative to each other on the vertical axis. Each light curve is binned at 5 minute intervals. The data from this work are displayed as dark and light grey dots for alternate light curves for clarity. The MOST data are displayed in black and the BSST data are displayed in light blue. The L2017 transit model is displayed as the light blue solid line for the BSST event while K2017 $\mathcal{M}_2$ transit models are displayed as the black solid lines for the MOST events.}
\label{fig:liu_monster}
\end{figure}

The three transit-like events connected by the L2017 ephemeris are not consistent with a strictly periodic signal, but can be described by a common ephemeris if TTVs on the order of $\sim20-40$ minutes are allowed. A series of transit events with TTVs on the order of half of the transit duration will ``smear out'' the events in a phased plot making them harder to detect. Therefore, to search for transit-like signals in our data that are consistent with the L2017 ephemeris plus TTVs, the phase-folded constituent light curves in Figure~\ref{fig:liu_fold} are shifted relative to each other on the vertical axis in Figure \ref{fig:liu_monster}. Each light curve has been binned at 5 minute intervals. The data from this work are displayed as dark and light grey dots for alternate light curves for clarity (since some light curves occasionally overlap in time). The MOST data are displayed in black and the BSST data are displayed in light blue. The transit models, phased to the L2017 linear ephemeris and shifted according to the TTV offsets listed in L2017 Table~2, are displayed as black and light blue solid lines for the K2017 $\mathcal{M}_2$ and L2017 models, respectively.

We first note that the ``20060605 RAE'' light curve shows a flux deficit with a time of event minimum that occurs $\sim1$ hour before the L2017 ephemeris predicted transit center time, and with flux deficit duration of $\sim1.5$ hours. However, the light curve shows a flux increase above the average value during the time of transit, which explains the slight increase in brightness during transit in Figure~\ref{fig:liu_fold}. Considering the higher points in the light curve to be the out-of-transit baseline, the flux deficit event is even deeper and longer in duration. Given the inconsistency of the flux deficit event with the L2017 transit model, and the additional variations in those light curves, we do not interpret it as being caused by a transiting exoplanet, and further, do not support the connection of the tentative L2017 event with the tentative MOST Signal C events though the TTV-based ephemeris.

The other light curves are relatively flat or contain variations that are not consistent with a transit event. Although the constituent light curves do not provide full phase coverage at each epoch, it seems unlikely that our data would have missed all transit events on the 15 epochs with partial light curve phase coverage. Our light curve on ``20140523 SS03'' was observed simultaneously with the ``20140523 MOST'' light curve. Unfortunately, the robotic telescope halted observations during part of the first of the two MOST Signal C events. In addition, the correct relative baseline of the data between -1 and 0 hours is unknown because of a large jump of the field on the detector at about -1 hours, and a median flip at about -0.5 hours. Therefore, despite our simultaneous observations, we cannot place strong constraints on the Signal C event. The two deeper events at -2.2 and +2.3 hours in the ``20140523 SS03'' light curve are analyzed in more detail in Section \ref{sec:discussion}. Since we are unable to predict the TTVs for our observed epochs, and since the SS03 robot shutdown during most of the Signal C event on UT 2014 May 23, our data are unable to completely rule out the L2017 reported TTV-based ephemeris. 

\subsection{Light Curves in Relation to Li2017 Ephemeris\label{sec:Li_reanalysis}}

Finally, we sought to phase-fold our light curve data according to the ephemeris proposed by Li2017. Unfortunately, due to the single transit-like event found by those authors, a precise period is not available for a phased transit search. The final observation from our campaign was obtained prior to the Li2017 reported event, so we have no simultaneous light curve to compare with theirs. 

However, Li2017 generously included full AIJ photometry measurements tables for all of their time-series observations on a public archive. With a measurements table loaded into AIJ, we were able to examine how the choice of different comparison star ensembles affected the Proxima light curve. 

We generally find that choosing comparison stars having brightness as close as possible to the target star reduces systematics in the data due to variable atmospheric conditions. In the Li2017 data, Proxima had an average of $\sim 1.5\times10^6$ net integrated counts in the aperture. Based on our re-analysis, it appears that Li2017 used comparison stars having $\sim 0.2\times10^6$ net integrated counts in the aperture, except for one that is about 50\% as bright as Proxima. There are three additional comparison stars that are more than $75\%$ as bright as Proxima, so we explored a re-reduction of the Li2017 photometry using a comparison ensemble which included only the four stars that are at least $50\%$ as bright as Proxima. 

\begin{figure}[!htb]
\centering \includegraphics[width=\linewidth, trim=0.0cm 0.0cm 0.0cm 1.2cm, clip=true]{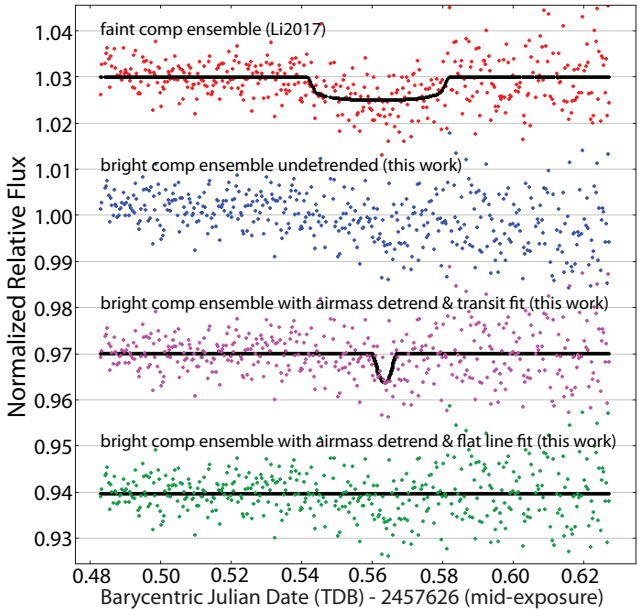}
\caption{Comparison of the Li2017 results with our alternate reduction. The original Li2017 light curve exhibiting the claimed transit is displayed at the top as red dots. The undetrended result using the bright star ensemble (see text) is shown as blue dots in the second light curve from the top. The simultaneously fitted and airmass-detrended bright ensemble light curve is shown as magenta dots in the third light curve from the top. The bright ensemble light curve simultaneously fitted to a flat line and airmass-detrended is shown as green dots at the bottom. Models are displayed as black solid lines. Light curves are successively shifted by 0.03 to minimize data overlap for clarity. The signal in the bright ensemble light curve is significantly shorter than the faint star ensemble used by Li2017.}
\label{fig:liComparison}
\end{figure}

The original Li2017 light curve exhibiting the claimed transit is displayed in Figure~\ref{fig:liComparison} as red dots and has been shifted on the vertical axis for clarity. The corresponding original transit model is displayed as the top black solid line. The undetrended result of using the bright star ensemble is displayed as blue dots. Notice a slight airmass trend downward on the right hand side and a significantly shorter event in the data. The simultaneously fitted and airmass-detrended light curve is displayed as magenta dots. The corresponding best fit model is represented by the middle black solid line. Finally, the same bright star result simultaneously fitted to a flat line and airmass-detrended is displayed as green dots. There is indeed a short residual event when fitting to a flat line. This could indicate that the short transit-like signal is a {\it bona fide} astrophysical event. On the other hand, since even the brightest comparison stars are still $\sim 25\%$ fainter than Proxima, the short signal could be a residual systematic, albeit much shorter than the signal resulting from the faint star ensemble. With the currently available data, we are unable to conclude which of the results best represents the true behavior of the Li2017 Proxima light curve on UT 2016 August 25. 

Li2017 also provided on the public archive AIJ photometric tables for 22 additional Proxima time-series observations. We re-investigated those 22 light curves and found seven light curves that show apparent events having various durations and depths, any of which could be a Proxima astrophysical event or a systematic (we did not investigate alternate comparison star ensembles for these observations). It is unlikely that seven out of 22 blind search observations would catch transit events, which suggests that the variations were caused by an alternate astrophysical mechanism, or were caused by systemics, or a combination of both. These non-periodic variations exhibit a range of behaviors similar to the behaviors observed in our data (see Figure~\ref{fig:snippet} and the Appendix\ref{sec:appendix}). 

\section{Discussion}\label{sec:discussion}

We find no compelling evidence for Proxima~b transits corresponding to any of the previously published ephemerides. We do, however, find many examples of light curves having variations consistent with the predicted $0.5-1.3\%$ Proxima~b transit depths. \citet{Davenport:2016} predicted that low energy flares of this magnitude occur approximately every 20 minutes on Proxima. These semi-regular events, having an amplitude similar to the predicted Proxima b transit depth (assuming Proxima b transits do exist), and occurring on the time scale of the predicted Proxima b transit duration, could contribute to the variations seen in our data. These positive-flux events  would bias the individual light curve normalization levels upward by varying amounts, depending on the amount of low energy flare activity within the time-period covered by a particular light curve. Positive flux events occurring during a {\it bona fide} transit would tend to obscure the transit by changing the apparent duration, shifting the apparent transit center time, and/or dividing the transit into two or more shorter events, significantly complicating the detection of a potential real Proxima~b transit. Furthermore, it is possible that starspots forming or changing significantly on $\sim\rm{hour}$ timescales could produce photometric dips similar to the transit-like events and other variations detected in this work and by other authors.

As mentioned in Section~\ref{sec:previous}, K2017 found only tentative evidence for Proxima~b transit events, and they too discuss the difficulties of detecting Proxima~b transits, if in fact they do occur, given the predicted low energy flare contributions to the Proxima light curve data. Because of the pervasiveness of variations in our data, we conclude that the low energy flares and starspot transients, combined with light curve systematics, are the source of many or all of the variations in our data. However, the variations could be hiding {\it bona fide} transit events. 

We reiterate that since we do not know the appropriate underlying astrophysical model that describes Proxima's light curve behavior, our light curve data presented here have been individually detrended assuming a flat light curve model. We have visually compared each undetrended light curve with its detrended version to verify that our detrending method did not significantly reduce or enhance transit-like events in our data.  

To further investigate systematics and other variations in ground-based Proxima light curves, we also re-analyzed the data set that included the Li2017 $2.5\sigma$ transit-like signal detection and find that the signal becomes less obvious when comparison stars closer in brightness to Proxima are used. In addition, we reviewed 22 additional light curves provided by Li2017 and found seven that show variations, any of which could be Proxima astrophysical events or systematics. It appears that the Li2017 observations exhibit a range of behaviors similar to the behaviors observed by us. 

\begin{figure*}[!htb]
\centering \includegraphics[width=\linewidth, trim=0.0cm 0.0cm 0.0cm 0.0cm, clip=true]{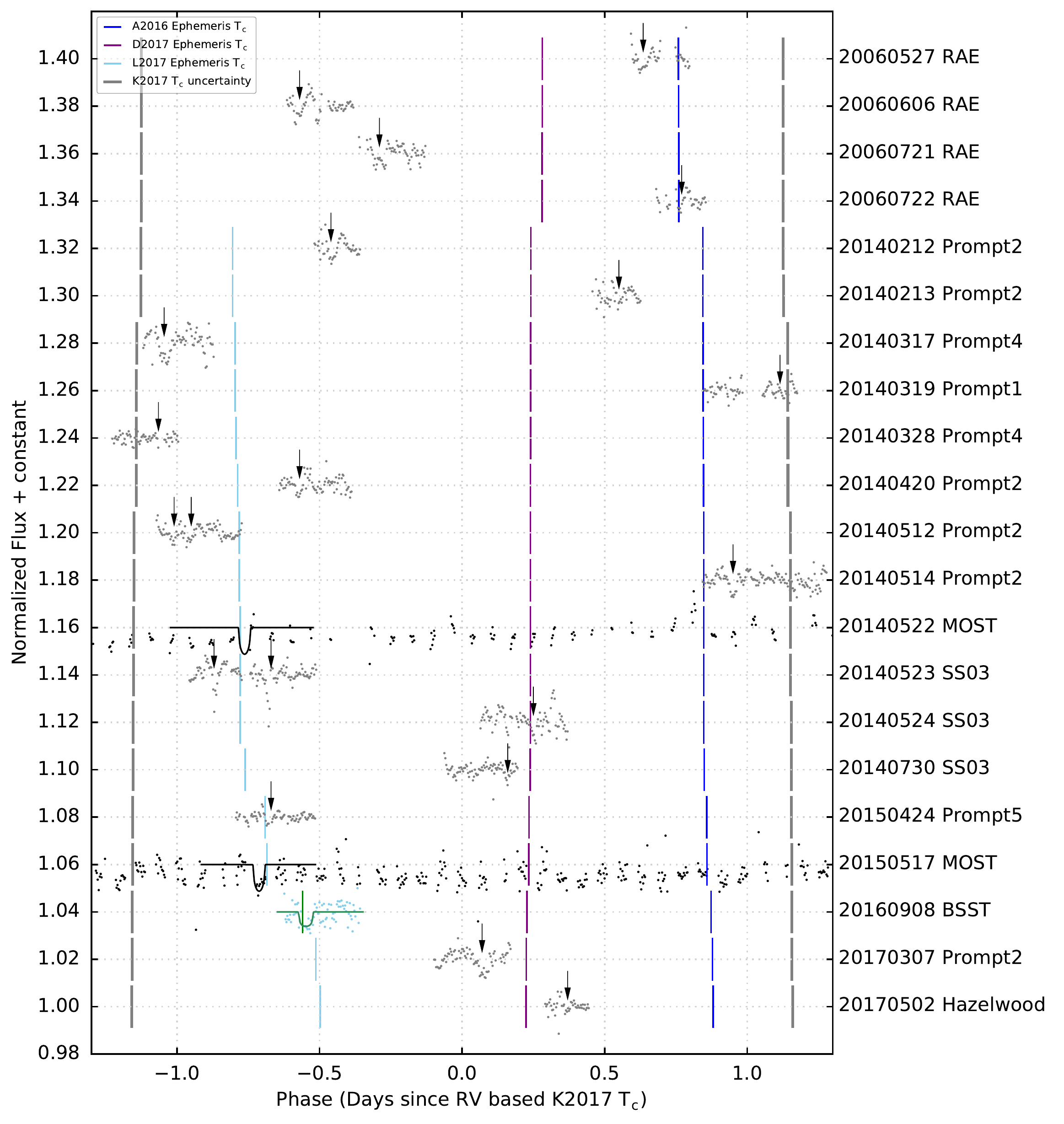}
\caption{A subset of 18 light curves from this work that display variations with an amplitude similar to the depth predicted for a transiting Proxima b planet. All light curves are binned in 5 minute intervals and phased to the K2017 RV-based ephemeris. Light curves from this work are displayed as grey dots. The MOST and BSST data are plotted as black and light blue dots, respectively, along with the K2017 $\mathcal{M}_2$ (Signal C) and L2017 transit models in black and green solid lines, respectively. The black arrows are placed to highlight events that exhibit variations similar to the claimed transit detection depths reported in Table \ref{tbl:litephemerides}. The transit center times corresponding to the ephemerides of A2016, D2017, and L2017 are plotted as vertical blue, magenta, and light blue bars, respectively. The gray vertical bars mark the $\pm2\sigma$ uncertainty boundaries for K2017 RV-based ephemeris.}
\label{fig:interesting}
\end{figure*}

As visual evidence of the routine variability that may mimic transit-like events, we present a collection of our Proxima light curve data within $\pm2\sigma$ of the K2017 RV-based ephemeris in Figure~\ref{fig:interesting}. We point out events with arrows that could be interpreted as transit-like features. In some cases, the events occur very near the time corresponding to one of the previously claimed transit ephemerides. However, such events also occur with similar frequency at other times. 

\begin{figure}
    \centering
    \includegraphics[width=\columnwidth, trim=0.0cm 0.0cm 0.5cm 0.0cm, clip=true]{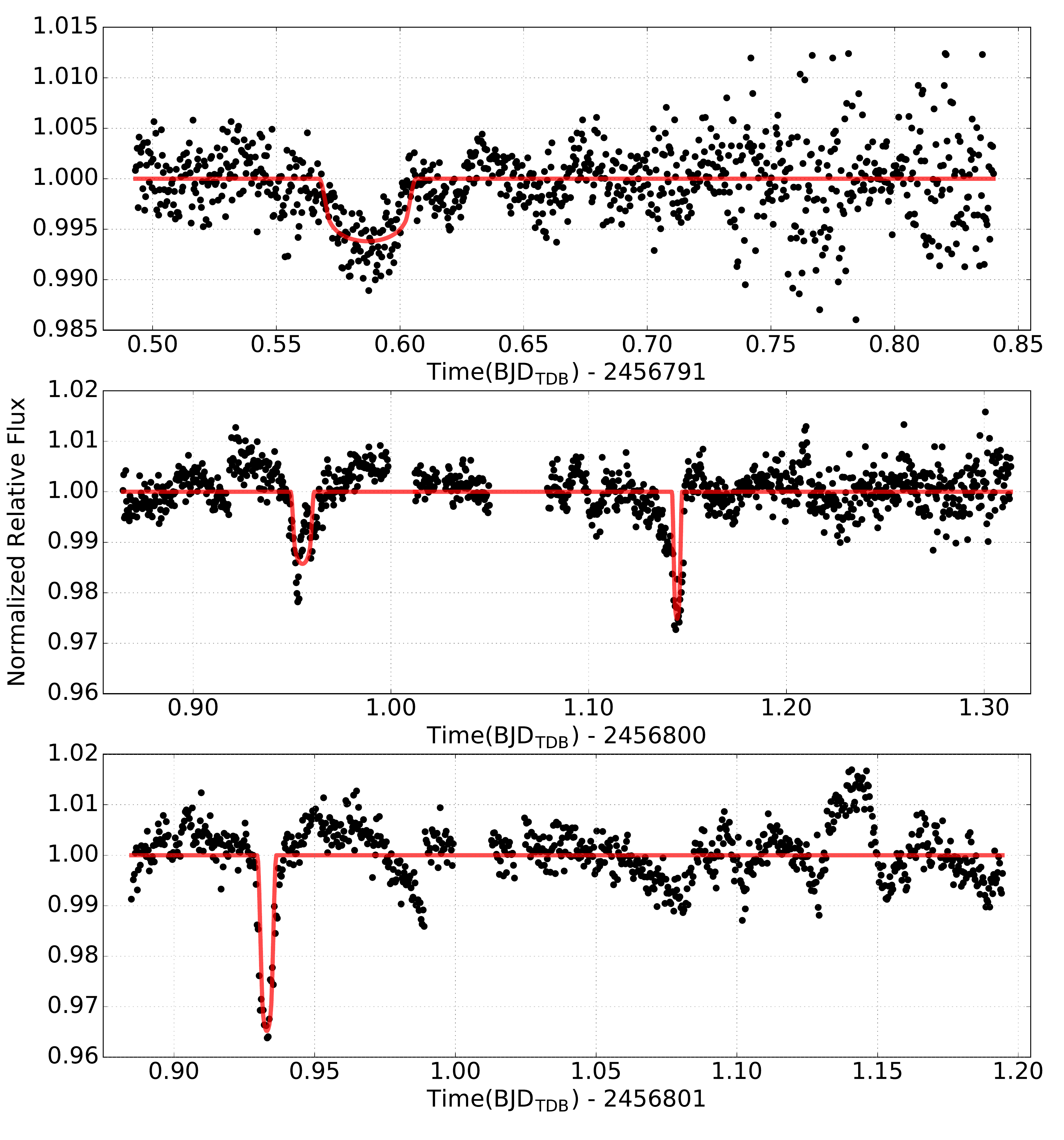}
    \caption{Examples of transit-like events in our Proxima data. The unbinned, detrended data are displayed as black dots, and the best fit transit models are displayed as red solid lines. (Top) Prompt 2 $R$ band light curve from UT 2014 May 14. The model has a duration of 56 minutes and a depth of $0.55\pm0.1\%$. The somewhat asymmetric morphology and the post-egress sawtooth-shaped variations suggest that this light curve feature may not have been caused by a transiting exoplanet. (Middle) Prompt SS03 $R$ band light curve from UT 2014 May 23. (Bottom) Prompt SS03 $R$ band light curve from UT 2014 May 24. It is unlikely that the short, deep, mostly asymmetric events in the middle and bottom panels were caused by transits of Proxima~b. See text for more details.}
    \label{fig:transit_like_event}
\end{figure}

In Figure~\ref{fig:transit_like_event}, we present a more detailed investigation of four of the transit-like signals identified in Figure~\ref{fig:interesting}. The unbinned, detrended data are shown as black dots, and the best fit \citet{Mandel:2002} transit models are shown as a solid red lines. The ``20140514 Prompt 2'' light curve is presented in the top panel of Figure~\ref{fig:transit_like_event}. The best fit model has a duration of 56 minutes and a depth of $0.55\pm0.1\%$, which is somewhat consistent with many of the tentative detections from the literature. However, the transit center time is not consistent with the other photometric-based ephemerides from the literature, and there are no other obvious events in our data that would indicate that the fitted event is periodic at the RV-based period. Those inconsistencies combined with the somewhat asymmetric morphology and the post-egress saw-tooth-shaped variations suggest that the $\sim5\sigma$ detection is unlikely to have been caused by a transiting exoplanet.

We also found three transit-like features that are deeper than predicted for Proxima~b. Two of the events occur in the ``20140523 Prompt SS03'' light curve shown in the middle panel of Figure~\ref{fig:transit_like_event}. The event centered at 2456800.956\,\bjdtdb\ has a best fit transit model depth of 1.31\% and a duration of $\sim 25$~minutes. However, the model fit does not find the correct pre- and post-transit baseline, and doesn't fully account for the very short, initially deeper, ingress feature. Accounting for both of those features, we estimate that the true maximum change in the light curve is $\sim 3$\%. The second event in the middle panel, centered at 2456801.145\,\bjdtdb, has a best fit transit model with a depth of 2.5\% and a duration of 11 minutes. Both events in this light curve have significantly asymmetric ingresses and egresses. Our ``20140524 Prompt SS03'' light curve has the deepest transit-like event found in our data and is displayed in the bottom panel of Figure~\ref{fig:transit_like_event}. The event is centered at 2456801.933\,\bjdtdb\, and the best fit transit model has a depth of 3.3\% and a duration of 12 minutes.

We have investigated all of our systematics indicators and find no parameters that are correlated with these transit-like signals. However, due to the robotic nature of the Prompt telescopes, and the limited set of systematics indicators available to us, we cannot exclude systematics as the source of these light curve features. In fact, Figure~\ref{fig:liu_monster} illustrates that the MOST observations provide partial coverage of the event centered at 2456800.956\,\bjdtdb\ and do not show evidence of an event at that time ($\sim -2.3$ hours in Figure \ref{fig:liu_monster}). If these events are in fact astrophysical in nature, it is unlikely that they were caused by a transit of Proxima~b due to asymmetric feature morphology, events that are too deep and short, and/or the fact that three similar features occurred within one day around 2456801\,\bjdtdb, but are not found elsewhere in our data.
\\
\\

\section{Conclusion}\label{sec:conclusion}

From a total of 262 Proxima light curves that will be published in Paper II of this series, we presented 96 Proxima time-series photometric observations that correspond to previously published Proxima~b ephemerides from the literature. The light curves span from 2006 to 2017 and were conducted using a combination of RAE, Skynet and KELT-FUN telescopes. Because almost all of our observations were conducted before the A2016 RV-discovered planet was announced, we were generally conducting a blind search for transits of Proxima. 

Although Proxima's flares are prominent across the UV and optical bands, they are indeed stronger in the blue compared to the underlying stellar photosphere \citep{Walker:1981}, so we targeted the $R$ passband and redder to minimize the impact on our photometric observations. Even so, contamination from systematics, starspot transients, flares, and/or low energy flares that are predicted to occur about every $\sim20$ minutes at the 0.5\% level \citep{Davenport:2016}, is significant in our photometric data.

We simultaneously cleaned, detrended, and normalized each night of differential photometry individually using a $3\sigma$ iterative cut and a flat light curve model in lieu of a correct, but unknown model. We have visually compared each pre-cleaned, undetrended light curve with its cleaned and detrended version to verify that the data processing methods we used did not significantly reduce or enhance transit-like events in our data.

We investigated our data in relation to the RV-based ephemerides presented in A2016, K2017, and D2017, and the photometric-based ephemerides presented in K2017, L2017, and Li2017. In general, we find pervasive variability in our cleaned and detrended light curve data at the level of $0.5-3.0\%$. We also explored a re-analysis of the Li2017 data using a different ensemble of comparison stars that were similar in brightness to Proxima and found a significant reduction in the duration of the claimed event (the event was essentially eliminated). We also find variability in seven of 22 additional Li2017 light curves, similar to what we find in our data. Overall, considering all of the available data that coincide specifically with the previously published claimed transit detections, we are unable to independently verify those claims. We do, however, verify the previously reported ubiquitous and complex variability of the host star. 

In Paper~II, we will present a search for periodic Proxima~b transits over the range of periods having good phase coverage from our full set of 262 light curves. We will also present an analysis of the transit detection sensitivity of our data across a range of transit model parameters. 

As previously mentioned, flares are stronger in blue bands than red bands, but are still significant in red bands. The Transiting Exoplanet Survey Satellite (TESS, \citealt{Ricker:2015}) will likely observe Proxima for at least 27 days. TESS observes in a single band that includes the optical wavelengths above $\sim600$\,nm ($\sim R$~band and redder), which will help minimize the contamination from Proxima's flares. However, it may still be difficult to separate contamination from the combination of possible starspot transients and the predicted every $\sim$20-min low energy flares from a potential {\it bona fide} transit signal in the TESS data, especially if the Proxima~b orbit is not strictly periodic due to significant perturbations from other companion(s) in the system. One potential approach to separate the predicted low energy flares from potential real transit signals would be to conduct simultaneous observations in a blue and red band, since the low energy flares should be more significant in the blue band, while the transit signal should be consistent in both. Alternatively, observations simultaneous with the TESS observations, but in a different filter band could help differentiate transit signals in the TESS data.

\acknowledgements
The authors thank the anonymous referee for helpful suggestions that improved this manuscript.
We also thank Mike Lund, Ryan Oelkers, and Rob Siverd for thoughtful discussions regarding our data reduction and analysis. We also thank the former director of the Perth Observatory, Jamie Biggs for the observing time on the RAE telescope and to Arrie Verveer for keeping the telescope working. We thank Greg Laughlin for his encouragement on this and related projects.

The RAE Perth Robotic Telescope is a collaboration between the Perth 
Observatory, Lawrence Berkeley National Laboratory, and the 
Hands-On-Universe project (HOU; Co-Directed by Carl Pennypacker and 
Alan Gould; http://lhs.berkeley.edu/hou) at Lawrence Hall of Science 
(LHS) of the University of California, Berkeley.  Establishment of 
the telescope was made possible by United States National Science 
Foundation (NSF) grant ESI 0125757: The Real Astronomy Experience 
(RAE), an Exhibit for ISE Centers (HOU Co-PIs Pennypacker and Gould).

Dax Feliz gratefully acknowledges the support
from NSF “Graduate Opportunities at Fisk in Astronomy and Astrophysics Research” (GO-FAAR) grant \# 1358862.

\appendix
\setcounter{secnumdepth}{0}
\section{Full light curve data}
\label{sec:appendix}

\begin{figure*}
    \centering
    \includegraphics[width=0.99\linewidth]{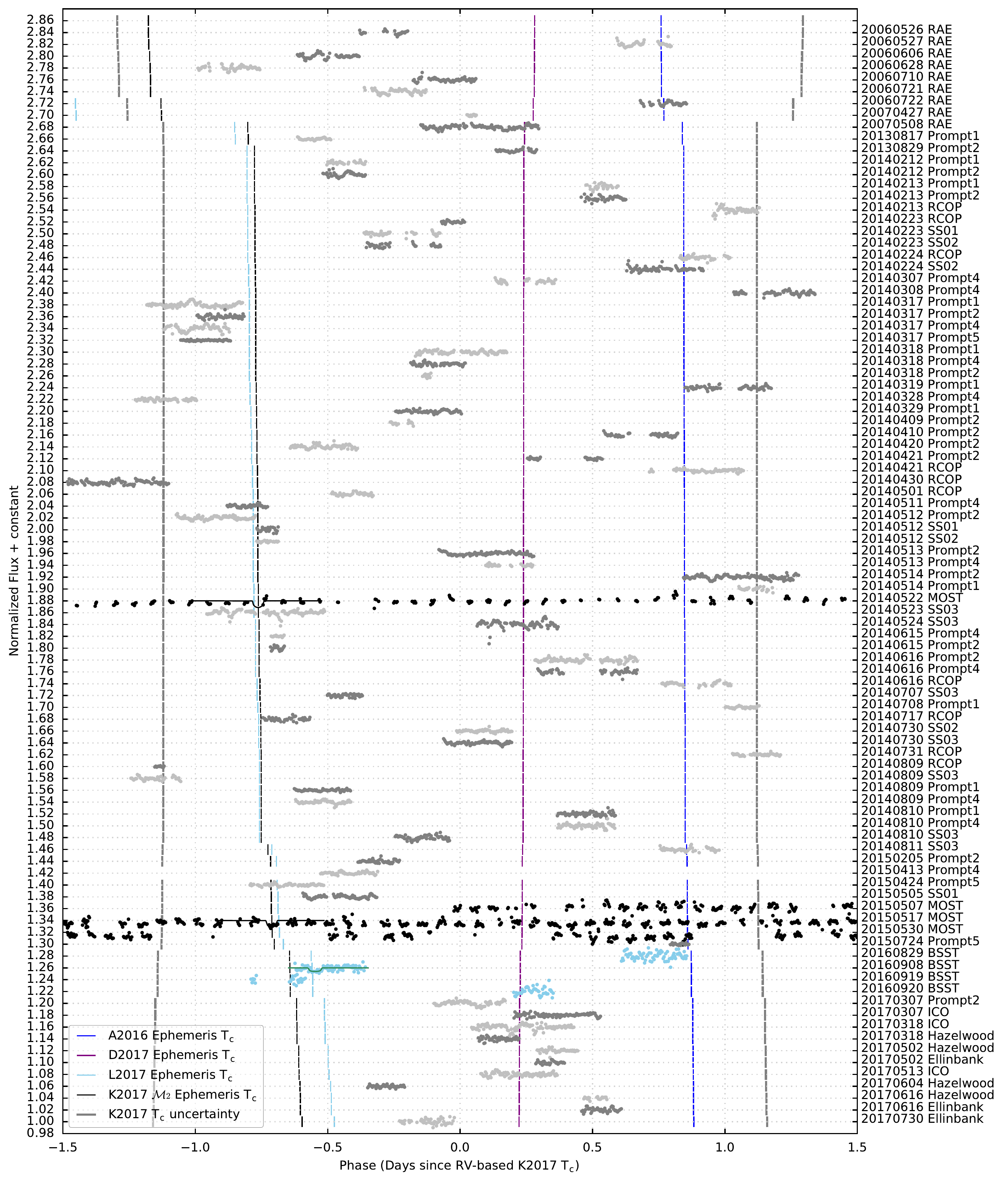}
    \caption{The full set of 85 light curves, along with the literature light curves, that contribute within $\pm2\sigma$ of the K2017 RV-based ephemeris. The grey vertical bars at $\sim\pm1.2$\,day mark the extents of the $2\sigma$ uncertainty. Each light curve is binned at 5 minute intervals. The data from this work are displayed as dark and light grey dots for alternate light curves for clarity when overlapping. The MOST data are displayed as black dots, and the BSST data are displayed as light blue dots. The K2017 $\mathcal{M}_2$ and L2017 transit models are displayed as black and green solid lines for the MOST and BSST events, respectively. The transit centers predicted by the A2016 (A2016) RV-based, D2017, K2017 Signal C, and L2017 ephemerides are shown as blue, magenta, black, and light blue vertical bars, respectively.}
    \label{fig:full_monster}
\end{figure*}

In Figure~\ref{fig:full_monster} we present all 85 of our new Proxima light curves, along with the literature light curves, that contribute within $\pm2\sigma$ of the K2017 RV-based ephemeris. The figure is best viewed electronically due to its large format. All light curves are folded on the K2017 RV-based ephemeris and shifted relative to each other on the vertical axis for clarity. The ``20060605 RAE'' light curve from Figure \ref{fig:liu_monster} does not fall within the phase range covered by Figure \ref{fig:full_monster}. If we extended our x-axis, this light curve would appear at $\sim-1.5$ days from K2017 RV-based ephemeris.

The grey vertical bars at $\sim\pm1.2$ days indicate the extents of the $2\sigma$ uncertainty, which is altogether $\sim2.5$ days, but varies depending on the amount of time since the reference epoch, T$_0$, due to the cumulative uncertainty in the period. Also shown are the transit centers at each displayed epoch, extracted from other literature ephemerides listed in Table \ref{tbl:litephemerides}, after phasing to the K2017 RV-based ephemeris. The transit centers predicted by the A2016 RV-based ephemeris are displayed as blue vertical lines. The transit centers predicted by the D2017 ephemeris are displayed as magenta vertical solid lines. The K2017 Signal C and L2017 ephemerides are shown as black and light blue vertical bars, respectively.

Each light curve is binned at 5 minute intervals. The data from this work are displayed as dark and light grey dots for alternate light curves for clarity when overlapping. The MOST data are displayed as black dots, and the BSST data are displayed as light blue dots. The L2017 transit models are displayed as black and green solid lines for the MOST and BSST events, respectively.

\begin{figure}
    \centering
    \includegraphics[width=0.7\linewidth]{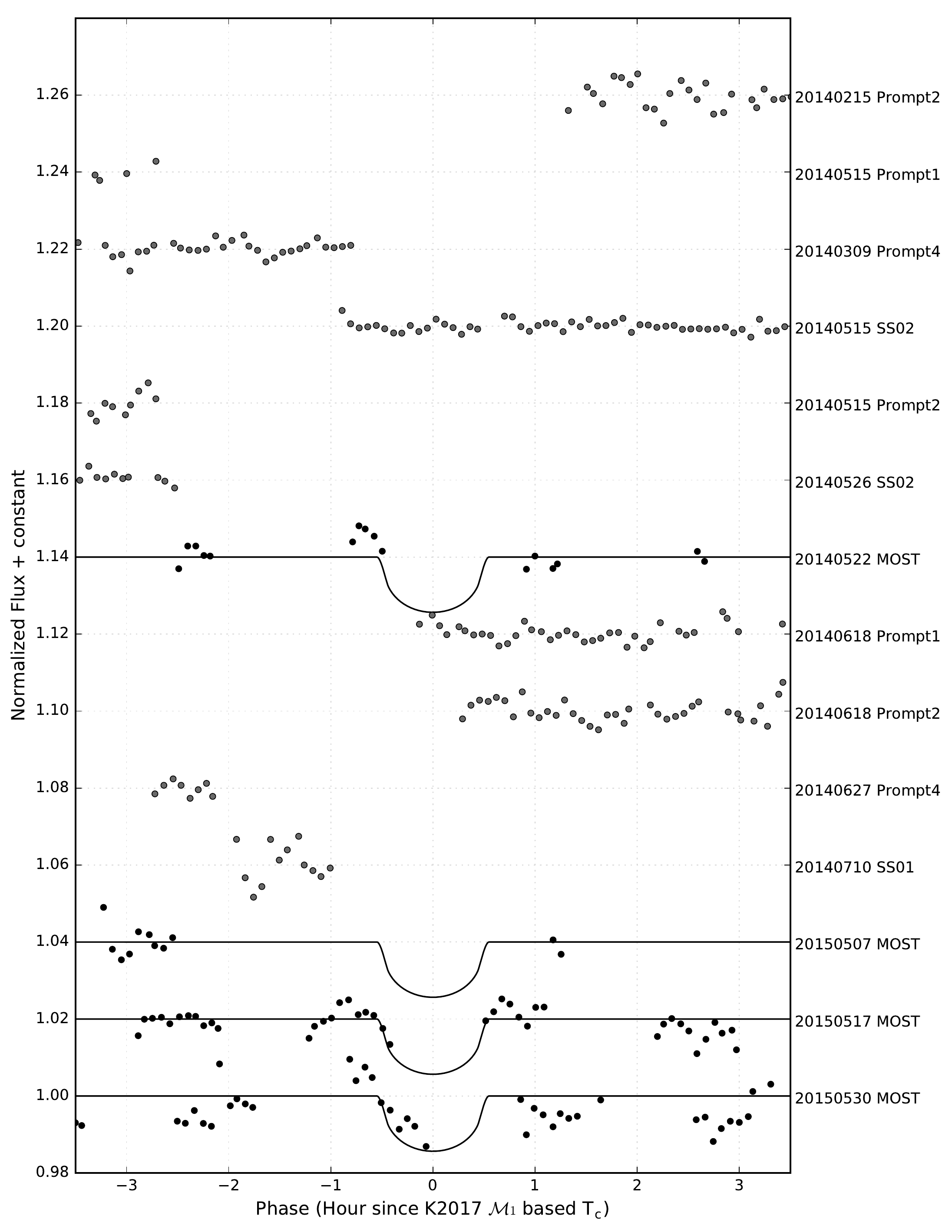}
    \caption{The full set of 10 light curves that contribute within $\pm 3$ hours of the K2017 Model $\mathcal{M}_1$ ephemeris. The light curves from this work are shown as grey dots. The MOST data are shown as black dots. The K2017 $\mathcal{M}_1$ transit models are over-plotted as black solid lines on the MOST light curves.}
    \label{fig:full_M1}
\end{figure}

The K2017 Model $\mathcal{M}_1$ ephemeris does not overlap within $\pm2\sigma$ of the K2017 RV-based ephemeris. In Figure \ref{fig:full_M1} we present our 10 light curves that contribute within $\pm 3$ hours of the K2017 Model $\mathcal{M}_1$ ephemeris. Each light curve is binned at 5 minute intervals. The light curves from this work are displayed as grey dots. The MOST data are displayed as black dots. The $\mathcal{M}_1$ transit models are over-plotted as black solid lines on the MOST light curves.  

All of our light curve data will be provided in machine readable format as part of Paper~II.

\bibliographystyle{apj}

\bibliography{output.bbl}

\end{document}